\documentclass[prd,nofootinbib,twocolumn, floatfix, superscriptaddress]{revtex4}
\usepackage[utf8]{inputenc}
\pdfoutput=1

\usepackage{adjustbox}
\usepackage{tabularx}
\usepackage{array} 
\usepackage{comment}
\usepackage{graphicx}

\usepackage{epsfig}
\usepackage{bm}
\usepackage{amssymb}
\usepackage{float}
\usepackage{amsmath}
\usepackage{dcolumn}
\usepackage{cancel}
\usepackage[colorlinks]{hyperref}
\usepackage[usenames, dvipsnames]{color}

\hypersetup{
     breaklinks=true, 
    pdfstartview={FitH},  
    colorlinks=true, 
    linkcolor=blue,  
    citecolor=red,  
    filecolor=magenta,  
    urlcolor=blue, 
    anchorcolor=green,  
    linktocpage=true
}

\providecommand{\U}[1]{\protect\rule{.1in}{.1in}}

\newcommand{\be}{\begin{equation}}
\newcommand{\ee}{\end{equation}}

\newcommand{\mincir}{\raise
-3.truept\hbox{\rlap{\hbox{$\sim$}}\raise4.truept\hbox{$<$}\ }}
\newcommand{\magcir}{\raise
-3.truept\hbox{\rlap{\hbox{$\sim$}}\raise4.truept\hbox{$>$}\ }}

\ifx\pdfoutput\relax\let\pdfoutput=\undefined\fi
\newcount\msipdfoutput
\ifx\pdfoutput\undefined\else
\ifcase\pdfoutput\else
\ifx\paperwidth\undefined\else
\ifdim\paperheight=0pt\relax\else\pdfpageheight\paperheight\fi
\ifdim\paperwidth=0pt\relax\else\pdfpagewidth\paperwidth\fi
\fi\fi\fi

\hypersetup{colorlinks=true,
	breaklinks=true,
	pdfstartview=Fit,
	linkcolor=blue,
	citecolor=blue,
	urlcolor=blue}

\begin{document}
\title{Baryogenesis constraints on generalized mass-to-horizon entropy}

\author{Giuseppe Gaetano Luciano}
\email{giuseppegaetano.luciano@udl.cat}
\affiliation{Departamento de Qu\'{\i}mica, F\'{\i}sica y Ciencias Ambientales y 
del Suelo,  Escuela Polit\'ecnica Superior -- Lleida, Universidad de Lleida, Av. 
Jaume II, 69, 25001 Lleida, Spain}

\author{Emmanuel N. Saridakis}
\email{msaridak@noa.gr}
\affiliation{National Observatory of Athens, Lofos Nymfon 11852, Greece}
\affiliation{CAS Key Laboratory for Research in Galaxies and Cosmology, School 
of Astronomy and Space Science, University of Science and Technology of China, 
Hefei 230026, China}
\affiliation{Departamento de Matem\`{a}ticas, Universidad Cat\`{o}lica del 
Norte, Avda.
Angamos 0610, Casilla 1280 Antofagasta, Chile}

\begin{abstract} 
We investigate the generation of the baryon asymmetry  within the   
cosmological framework based on a generalized mass-to-horizon entropy. This 
entropy, recently proposed as a power-law extension of the Bekenstein-Hawking 
area law, arises from a modified mass-horizon relation constructed to ensure 
consistency with the Clausius relation. By applying the gravity-thermodynamics 
conjecture, the resulting corrections to the Friedmann equations modify the 
evolution of the Hubble parameter. Consequently, even the standard supergravity 
coupling between the Ricci scalar and the baryon current can generate a 
non-vanishing matter-antimatter asymmetry. Comparison with observational data 
yields a stringent constraint on the entropic exponent, namely $0 < 1 - n 
\lesssim \mathcal{O}(10^{-2})$, at the decoupling temperature 
$T_D \simeq 10^{16}\,\text{GeV}$, 
corresponding to the current upper limit on tensor-mode fluctuations at the 
inflationary scale. These findings indicate that minor, subtle, yet physically 
significant departures, from the standard Bekenstein-Hawking entropy ($n = 1$) 
may be required to achieve full consistency with present cosmological 
observations.
\end{abstract}

\maketitle

\section{Introduction}
\label{SecI}

Recent cosmological measurements, such as those of supernova luminosity 
distances \cite{SupernovaSearchTeam:1998fmf,SupernovaCosmologyProject:1998vns}, 
anisotropies in the cosmic microwave background \cite{COBE:1992syq,WMAP:2003ivt} 
and large-scale structure surveys 
\cite{2DFGRS:2001zay,SDSS:2003eyi,BOSS:2016wmc}, strongly indicate that the 
Universe has experienced two separate periods of accelerated expansion: an 
initial inflationary epoch and the present-day cosmic acceleration. 
These findings have, in turn, inspired two main theoretical approaches to 
account for such observations.

One line of research focuses on modifying the underlying geometric framework of 
General Relativity (GR). Rather than strictly following Einstein’s original 
formulation, extensions of the Einstein-Hilbert action are explored, giving rise 
to a wide class of models collectively referred to as modified gravity 
theories~\cite{Capozziello:2011et,CANTATA:2021asi}. 
A conceptually distinct direction maintains GR as the fundamental theory of 
gravity but introduces modifications within the matter sector. In this scenario, 
new dynamical ingredients, such as scalar fields (e.g., the inflaton) or dark 
energy fluids, are imposed as the driving sources of cosmic 
acceleration~\cite{Olive:1989nu,Bartolo:2004if,Copeland:2006wr,Cai:2009zp,
CosmoVerse:2025txj}.

Beyond these two main frameworks, a further independent line of thought has 
gained considerable attention, proposing a profound link between gravitational 
dynamics and 
thermodynamics~\cite{Jacobson:1995ab,Padmanabhan:2003gd,Padmanabhan:2009vy}. In 
this context, the Universe is treated as a thermodynamic system enclosed by the 
apparent horizon, and the gravitational field equations can be reproduced by 
applying the first law of thermodynamics to this 
boundary~\cite{Frolov:2002va,Cai:2005ra,Akbar:2006kj,Cai:2006rs}. 
Interestingly, this thermodynamic formulation is applicable not only to GR but 
also to a wide spectrum of modified gravity 
theories~\cite{Paranjape:2006ca,Akbar:2006er,Jamil:2009eb,Cai:2009ph}).  

Building on this thermodynamic perspective, the notion of horizon entropy 
becomes particularly relevant in the framework of entropic 
cosmology~\cite{Easson:2010av}. In this approach, thermodynamic considerations 
are employed more directly to model the large-scale evolution of the Universe, 
where horizon entropy gives rise to effective ``entropic forces'' that 
affect the cosmological dynamics. These corrections, motivated by boundary 
terms in the Einstein-Hilbert action~\cite{Easson:2010av}, have been proposed as 
a viable mechanism to account for the observed late-time accelerated expansion 
of the Universe.  

In light of the central role played by horizon entropy, it is natural to 
investigate the fundamental nature of entropy itself within the thermodynamic 
description of gravity. Understanding how gravitational dynamics emerge from 
microscopic degrees of freedom crucially depends on the underlying entropy-area 
relation. Therefore, particular attention has been devoted to exploring possible 
generalizations of the standard Bekenstein-Hawking entropy, which may involve 
quantum, statistical or geometric corrections and thus provide more information 
on the  emergent thermodynamic origin of spacetime.

Notable examples of such generalized entropy formulations include 
Rényi~\cite{renyi1961entropy}, Tsallis~\cite{Tsallis:1987eu,Tsallis:2009} and 
Sharma-Mittal~\cite{Sharma1975} entropies, which relax the assumption of 
additivity underlying the Boltzmann-Gibbs framework; Kaniadakis 
entropy~\cite{kaniadakis2001non,Kaniadakis:2002zz,Luciano:2024bco}, rooted in 
relativistic statistical mechanics; and Barrow entropy~\cite{Barrow:2020tzx}, 
motivated by quantum-gravitational corrections to horizon geometry 
(see~\cite{hanel2011comprehensive} for an axiomatic derivation of these 
generalized entropies). All of these formulations reduce to the classical 
Bekenstein-Hawking entropy in appropriate parameter limits, and their 
implications have been widely investigated 
in~\cite{Lymperis:2018iuz,Saridakis:2020lrg,Nojiri:2019skr,
Hernandez-Almada:2021rjs,Dheepika:2022sio,Jizba:2022icu,Lambiase:2023ryq,
Jizba:2024klq,Ebrahimi:2024zrk,Nojiri:2025gkq}.

Nevertheless, an important question arises about the validity of the 
thermodynamic description of gravity in the presence of these entropic 
deformations~\cite{Nojiri:2021czz,Gohar:2023lta}, namely whether it is 
theoretically consistent to modify the entropy while keeping the other 
thermodynamic quantities unchanged.
Several studies suggest that, due to the first law of thermodynamics, any 
modification of the entropy should be accompanied by corresponding adjustments 
in either the temperature or the internal energy of the 
system~\cite{Nojiri:2022sfd,Nojiri:2021czz}. 

Another perspective motivated by cosmological 
considerations~\cite{Gohar:2023hnb,Gohar:2023lta} arises from the observation 
that, provided the Clausius relation is employed to ensure thermodynamic 
consistency (i.e., to define the appropriate horizon temperature) and a linear 
mass-to-horizon relation (MHR) is assumed, entropic-force models become 
effectively indistinguishable from the standard framework based on Bekenstein 
entropy and Hawking temperature. This equivalence holds irrespective of the 
specific entropy function adopted on the cosmological horizon. As a result, all 
entropic cosmological scenarios constructed under these assumptions inevitably 
inherit the same shortcomings as the Bekenstein-Hawking approach, most notably 
its inability to account for the observed cosmological dynamics at both the 
background and perturbative levels~\cite{Basilakos:2012ra,Basilakos:2014tha}. To 
overcome this difficulty, a generalized MHR has been 
proposed~\cite{Gohar:2023hnb,Gohar:2023lta}, leading to a power-law modification 
of the entropy expression that encompasses, as particular cases, the 
Tsallis-Cirto~\cite{Tsallis:2013}, Barrow, and other non-standard entropy forms.

The cosmological consequences of the generalized mass-to-horizon entropy 
framework were recently investigated in Ref.~\cite{Gohar:2023lta}, showing that, 
for appropriate choices of the model parameters, the predictions are in good 
agreement with observational data. Furthermore, employing the 
gravity-thermodynamics correspondence, Ref.~\cite{Basilakos:2025wwu} derived 
modified Friedmann equations in which the additional contributions from the 
generalized entropy manifest as an effective dark energy sector. The 
corresponding dark energy equation-of-state parameter evolves dynamically, 
resembling either quintessence or phantom behavior at different redshifts, 
depending on the values of the entropic parameters. The resulting cosmological 
framework has also been demonstrated to remain consistent with current 
astrophysical bounds from baryon acoustic oscillations, including the recent 
DESI~DR2 survey~\cite{Luciano:2025ovj} (see also 
\cite{Ormondroyd:2025iaf,You:2025uon,Gu:2025xie,Santos:2025wiv,Li:2025cxn,
Carloni:2024zpl,Luciano:2025elo,Luciano:2025hjn,Chaussidon:2025npr,
Anchordoqui:2025fgz,Ye:2025ulq,Paliathanasis:2025dcr,Yang:2025mws,
Paliathanasis:2025hjw,Tyagi:2025zov} for further recent cosmological studies 
based on DESI data).

On the other hand, one of the long-standing open problems in modern cosmology 
concerns the origin of the baryon asymmetry of the Universe (BAU), for which a 
variety of theoretical approaches have been proposed over the 
years~\cite{Kolb:1983ni,Shaposhnikov:1987tw,Sakharov:1967dj}. A key 
interpretative framework was introduced by Sakharov~\cite{Sakharov:1967dj}, who 
first identified three necessary conditions that any CPT-invariant theory must 
satisfy in order to dynamically generate a non-vanishing asymmetry: \emph{(i)} 
violation of baryon number, \emph{(ii)} violation of charge conjugation (C) and 
charge-parity (CP) symmetries, and \emph{(iii)} departure from thermal 
equilibrium. However, in certain scenarios these requirements can be 
relaxed~\cite{Dolgov:1991fr}. For example, if CPT symmetry is dynamically 
broken~\cite{Cohen:1987vi}, a net BAU may arise even in the presence of thermal 
equilibrium. This idea underlies the mechanism of gravitational 
baryogenesis~\cite{Davoudiasl:2004gf}, where the baryon or lepton current is 
coupled to the derivative of the Ricci scalar, thus providing a natural source 
of matter-antimatter asymmetry.  Some
applications of gravitational baryogenesis can be found in 
Refs.~\cite{Davoudiasl:2013pda,Lambiase:2006dq,Fukushima:2016wyz,
Oikonomou:2016jjh, 
Odintsov:2016hgc,Ramos:2017cot,Luciano:2022pzg,Luciano:2022ely,Das:2021nbq, 
Troisi:2025ksj}.

Starting from these premises, this work investigates baryogenesis within the 
framework of cosmology based on generalized mass-to-horizon entropy.  
This formalism modifies the Friedmann dynamics and consequently introduces 
corrections to the Universe  energy density and pressure, allowing the 
generation and survival of a net baryon asymmetry.  

The structure of this work is as follows: in Sec.~\ref{SecII}, we implement the 
gravity-thermodynamics conjecture within the framework of 
generalized cosmology and derive the corresponding Friedmann equations. In 
Sec.~\ref{SecIII} we analyze the baryogenesis mechanism and obtain constraints 
on the entropic exponent through comparison with observations. Finally, 
Sec.~\ref{Conc} presents our conclusions and outlook. Unless explicitly stated otherwise, throughout the paper we adopt natural units.

\section{Modified Cosmology through generalized mass-to-horizon entropy}
\label{SecII}

We begin our analysis with a brief review of the gravity-thermodynamics 
conjecture in the framework of general relativity.
This analysis will be then generalized by incorporating the MHR together 
with the corresponding modified entropy proposed in 
\cite{Gohar:2023hnb,Gohar:2023lta}.  
For this purpose, our approach follows the methodology outlined in 
\cite{Basilakos:2025wwu}.

We conduct our analysis within a spatially flat 
Friedmann-Lema\^{\i}tre-Robertson-Walker (FLRW) background, described by the 
metric  
\be
ds^2 = g_{\mu\nu}dx^{\mu}dx^{\nu} = \ell_{\alpha\beta}dx^\alpha dx^\beta + 
\tilde r^2\left(d\theta^2 + \sin^2\theta\, d\phi^2\right)\, ,
\label{FRW}
\ee
where $\tilde r = a(t)\,r$, $x^0 = t$, $x^1 = r$, $\ell_{\alpha\beta} = 
\mathrm{diag}(-1,a^2)$, and $a(t)$ denotes the time-dependent scale factor.  
In addition, we assume that the Universe is filled with a perfect fluid of 
density $\rho$ and pressure $p$, respectively. The corresponding energy-momentum 
tensor takes the standard form, $T_{\mu\nu} = (\rho + p)\, u_{\mu} u_{\nu} + p\, 
g_{\mu\nu}$, where \( u^\mu \) denotes the four-velocity of the fluid. 
A further condition is provided by the energy-momentum conservation, $\nabla_\mu 
T^{\mu\nu} = 0$, which leads to the continuity equation
\begin{equation}
\label{cont}
\dot{\rho} + 3H(\rho + p) = 0\,.
\end{equation}
The associated work density, arising from variations of the apparent horizon 
radius, is defined as  
$\mathcal{W} = -\text{Tr}(T^{\mu\nu})/2 = (\rho - p)/2$, where the trace is 
taken with respect to the induced metric on the $(t,r)$ submanifold.  

Within this framework, the dynamical apparent horizon plays a key role in 
defining thermodynamic quantities.  For a spatially flat FLRW Universe, its 
radius is $\tilde r_A = 1/H$~\cite{Frolov:2002va,Cai:2005ra,Cai:2009qf}, where 
the Hubble parameter $H = \dot{a}/a$ characterizes the cosmic expansion rate 
(with the dot denoting differentiation with respect to time).  
The apparent horizon is assigned a Hawking-like 
temperature~\cite{Hawking:1975vcx}, namely 
\begin{equation}
\label{temp}
 T_h=-\frac{1}{2 \pi \tilde r_A}\left(1-\frac{\dot {\tilde
r}_A}{2H\tilde r_A}\right),
\end{equation} 
in analogy with black hole thermodynamics~\cite{Cai:2009qf,Padmanabhan:2009vy}. 
We further assume a quasi-static cosmological evolution, ensuring a well-defined 
horizon temperature at all times.  
The cosmic fluid is further taken to be in thermal equilibrium with the apparent 
horizon, consistent with long-term interaction 
mechanisms~\cite{Padmanabhan:2009vy,Frolov:2002va,Cai:2005ra,Izquierdo:2005ku,
Akbar:2006kj}.  
This assumption allows the use of standard thermodynamic laws without requiring 
non-equilibrium formalisms.  

The next step is to assign an entropy to the apparent horizon.  Within general 
relativity, this is traditionally done using the Bekenstein-Hawking formula, 
first introduced in black hole thermodynamics, $S_{BH} = A/(4G)$
where $A = 4\pi \tilde{r}_A^2$ denotes the area of the apparent 
horizon~\cite{Bekenstein:1973ur}.  

The gravity-thermodynamics conjecture states that Einstein’s field equations may 
be understood as emergent relations arising from local thermodynamic identities 
on causal horizons.  
In a cosmological framework, this interpretation implies that the Friedmann 
equations can be derived through the application of the first law of 
thermodynamics to the apparent horizon. 

To formalize this connection, we consider the thermodynamic relation  
\begin{equation}
dE = T_h\, dS + \mathcal{W}\, dV\,,
\label{14c}
\end{equation}
where \( dE \) is the infinitesimal change of the total energy $E=\rho V$ inside 
the apparent horizon during an interval \( dt \), as a consequence of the change 
in the volume \( dV = 4\pi \tilde{r}_A^2\, d\tilde{r}_A \). Using the 
definition~\eqref{temp} of the horizon temperature, Eq.~\eqref{14c} can be 
rewritten to yield the second Friedmann 
equation~\cite{Padmanabhan:2003gd,Padmanabhan:2009vy},
\begin{equation}
\dot{H} = -4\pi G\left(\rho + p\right)\,.
\label{F1}
\end{equation}
Hence, by inserting the continuity equation~\eqref{cont} and integrating both 
sides, one obtains the first Friedmann equation,  
\begin{equation}
    H^2 = \frac{8\pi G }{3}\rho + \frac{\Lambda}{3}\,,
    \label{F2}
\end{equation}
where the integration constant $\Lambda$ can be naturally identified with the 
cosmological constant.  
Since in the following we focus on the case of a radiation-dominated Universe, 
this contribution can be safely neglected.

\subsection{Modified Cosmology}
\label{MC}

As discussed in the Introduction, a generalized mass-to-horizon relation (MHR) 
was recently proposed in~\cite{Gohar:2023hnb,Gohar:2023lta} in the form
$M = \gamma \frac{c^2}{G} L^n$,  where \( M \) denotes the effective mass of the 
system, \( L \) is the cosmological horizon, \( \gamma \) is a constant with 
dimensions \( [L]^{1-n} \), and \( n \) is a non-negative real parameter. The 
speed of light \( c \) has been reinstated here for consistency with the 
conventions of~\cite{Gohar:2023hnb,Gohar:2023lta}.

By applying the Clausius relation and making use of the Hawking temperature 
$T_h$, one obtains the following generalized entropy 
formula~\cite{Gohar:2023hnb,Gohar:2023lta}:  
\begin{equation}
\label{GMHE}
    S = \gamma \,\frac{2n}{n+1}\, \tilde{r}_A^{\,n-1}\, S_{BH}\,,
\end{equation}
where \( S_{BH} \) corresponds to the usual Bekenstein--Hawking entropy, and the 
apparent horizon \( \tilde{r}_A \) serves as the characteristic length scale \( 
L \).

Let us clarify the physical significance of the parameters \( n \) 
and \( \gamma \), and   point out the limiting cases that connect this 
framework with known gravitational and cosmological models. In 
expression \eqref{GMHE}, the parameter \( n \) quantifies the effective 
dimensionality of the horizon degrees of freedom: for \( n>1 \) the entropy 
exhibits a super-extensive behavior, increasing more rapidly than the standard 
Bekenstein--Hawking area law, whereas \( n<1 \) corresponds to a sub-extensive 
regime with suppressed entropy growth. Moreover, the parameter \( \gamma \), 
on the other hand, functions as a fundamental normalization constant, setting 
the scale that links the microscopic informational content of the horizon to 
its macroscopic entropic representation~\cite{Gohar:2023hnb,Gohar:2023lta}.

Several notable limits of the entropy expression~\eqref{GMHE} shows its 
connection to established scenarios. For instance, for \( n=3 \)  the entropy 
scales as \( S\propto L^{4} \), while the corresponding mass grows with the 
enclosed volume, \( M\propto L^{3} \). The case \( n=2 \) leads to a mass 
proportional to the horizon area, \( M\propto L^{2} \), with the entropy 
adopting an extensive three-dimensional form, namely \( S\propto L^{3} 
\)~\cite{Gohar:2023hnb,Gohar:2023lta}. In the special limit \( n=1 \) and \( 
\gamma=1/2 \), one recovers the standard Misner--Sharp mass in spherical 
symmetry~\cite{Gong:2007md}, while choosing \( n=\gamma=1 \) reproduces the 
usual Bekenstein--Hawking area law. Since viable deviations from this scaling 
are expected to be small, in the subsequent analysis we focus on perturbative 
deviations around \( n=1 \), an assumption consistent with the observational 
bounds reported in~\cite{Gohar:2023lta,Basilakos:2025wwu,Luciano:2025ovj}. 
Furthermore, we set \( \gamma=1 \) (in units of $8\pi G=1$), 
following the treatment in~\cite{Basilakos:2025wwu}. From a theoretical 
perspective, this choice is reasonably justified, as \( \gamma \) enters 
\eqref{GMHE} merely as a multiplicative factor, so that any departure from unity 
is expected to be negligible or, at most, subdominant when compared with the 
effects induced by variations of the exponent \( n \).
On the phenomenological side, the assumption is further reinforced by recent 
observational analyses~\cite{Luciano:2025ovj}, which constrain \( \gamma \) to 
values very close to unity.

The generalized mass-to-horizon entropy relation \eqref{GMHE} was employed in 
Ref.~\cite{Basilakos:2025wwu} as the basis for constructing a modified 
cosmological framework. Specifically, invoking the gravity-thermodynamics 
conjecture and proceeding along the steps described above leads to the modified 
Friedmann equations
\begin{eqnarray}
\label{FM1}
    H^2&=&\frac{8\pi G}{3}\left(\rho+\rho_{DE}\right),\\[2mm]
    \dot H&=&-4\pi G \left(\rho+p+\rho_{DE}+p_{DE}\right),
    \label{FM2}
\end{eqnarray}
where the influence of the generalized entropy appears through an emergent 
effective dark energy component, whose energy density and pressure are given by
\begin{eqnarray}
\label{rhode}
    \rho_{DE}&\hspace{-1mm}=\hspace{-1mm}&\frac{3}{8\pi G}\left(H^2-\frac{2
n \gamma}{3-n}\hspace{0.2mm}H^{3-n}\right)\,,\\[2mm]
    \nonumber
    p_{DE}&\hspace{-1mm}=\hspace{-1mm}&\frac{1}{8\pi G}\left[
    2 n\gamma H^{1-n}\left(\dot H+\frac{3}{3-n}H^2\right)-\left(2\dot H+3H^2\right)
    \right].\\
    \label{pde}
\end{eqnarray}
It is easy to check that the standard Friedmann equations are recovered in the 
limit $n=1$, where $\rho_{DE}=p_{DE}=0$.

Substituting Eqs.~\eqref{rhode},\eqref{pde} into~\eqref{FM1},\eqref{FM2}, one 
obtains explicit expressions for the Hubble parameter and its time derivative, 
namely
\begin{eqnarray}
\label{MFr1}
    H &=& \left[\frac{4\pi 
G\left(3-n\right)\rho}{3n\gamma}\right]^{\frac{1}{3-n}}\,,\\[2mm]
    \dot H &=&-\frac{4\pi G H^{n-1}}{n\gamma}\left(\rho+p\right).
    \label{MFr2}
\end{eqnarray}
These equations constitute the starting point for investigating the baryogenesis 
mechanism within the present framework. In particular, they encode the 
modifications to the background cosmological dynamics induced by the generalized 
entropy formalism, and therefore provide the necessary input for evaluating how 
such deviations may affect the generation of the observed matter-antimatter 
asymmetry in the early Universe.

\section{Baryogenesis}
\label{SecIII}

The origin of the matter-antimatter asymmetry in the early Universe remains one 
of the most fundamental open problems in modern cosmology. Observations clearly 
demonstrate that matter dominates over antimatter, in contrast with the 
expectations of the Standard Model of particle physics~\cite{Canetti:2012zc}. 

Among the various models proposed for baryogenesis, supergravity (SUGRA) 
frameworks 
can offer a viable mechanism for producing a net baryon asymmetry in the early 
Universe~\cite{Kugo:1982mr}. Within this context, the (dynamical) CPT violation 
arises through an 
interaction that couples the derivative of the Ricci scalar curvature, 
$\partial_{\mu}R$, 
to the baryon/lepton current $J^{\mu}$ \cite{Davoudiasl:2004gf}, i.e.
\begin{equation}
\label{Jmu}
\mathcal{S}_{\text{int}}=\frac{1}{M_*^2}\int 
d^4x\sqrt{-g}\,J^{\mu}\partial_{\mu}\mathcal{R}\,,
\end{equation}
where $M_*$ denotes the cutoff mass scale characterizing the effective theory 
and is typically taken to be at the Planck scale, i.e.\ $M_* = (8\pi 
G)^{-1/2}$~\cite{Bento:2005xk,Sadjadi:2007dx,Das:2021nbq,Luciano:2022pzg}.

In an expanding cosmological background where
$\mathcal{R}$ evolves with time, the derivative $\partial_{\mu}\mathcal{R}$
acts as a classical, non-vanishing field which differentiates between matter
and antimatter.  This induces an effective, time-dependent chemical potential
for baryons and antibaryons and drives a net number density even in thermal
equilibrium.  Once baryon-violating processes become inefficient at a
characteristic decoupling temperature, the produced asymmetry is frozen and
persists thereafter.  In this way the curvature-current coupling supplies the
necessary ingredient for baryogenesis without introducing extra scalar
degrees of freedom.

From the viewpoint of Sakharov’s criteria, this mechanism realises a
generalized version of the three conditions.  Baryon (or lepton) number
violation originates from high-energy interactions already present in the
underlying theory; the coupling to $\partial_{\mu}\mathcal{R}$ effectively
violates $C$ and $CP$ by shifting particle and antiparticle energies; and the
role of departure from equilibrium is played by the  curvature
background, which breaks $CPT$ spontaneously while the plasma remains thermal.  
Thus, the operator in Eq.~\eqref{Jmu} provides a concrete and minimal way to
generate the observed baryon asymmetry within supergravity and related
frameworks. 

To quantify the asymmetry generated by the coupling~\eqref{Jmu}, let us observe 
that 
in the case of an expanding Universe filled with a perfect fluid characterized 
by the four-velocity $u^{\mu}$, one has~\cite{Kugo:1982mr}
\begin{equation}
    J^{\mu}=\left(n_B-n_{\bar B}\right)u^\mu\,,
\end{equation}
where $n_B$ and $n_{\bar B}$ denote baryon and anti-baryon number density, 
respectively. Specialising to the comoving frame of the fluid, where 
$u^\mu=(1,0,0,0)$,
it follows that $J^\mu=\bigl(n_B-n_{\bar B},\,0,0,0\bigr)$.

Furthermore, in a spatially homogeneous FRW background, the Ricci scalar depends 
only on cosmic time,
$\mathcal{R}=\mathcal{R}(t)$, so that
\begin{equation}
\partial_\mu\mathcal R=(\dot{\mathcal R},\,0,0,0)\,.
\end{equation}
Accordingly, the contraction appearing in Eq.~\eqref{Jmu} reduces to 
$J^\mu\partial_\mu\mathcal R=\bigl(n_B-n_{\bar B}\bigr)\,\dot{\mathcal R}$, and 
the interaction density becomes
\begin{equation}
\mathcal{L}_{\text{int}}=\frac{1}{M_*^2}\,J^\mu\partial_\mu\mathcal R
=\frac{\dot{\mathcal R}}{M_*^2}\,\bigl(n_B-n_{\bar B}\bigr)\,.
\end{equation}
It is then convenient to define the effective baryonic chemical 
potential~\cite{Davoudiasl:2004gf}
\begin{equation}
\label{pot}
\mu_B(t)=-\mu_{\bar  B}(t)\equiv-\frac{\dot{\mathcal R}(t)}{M_*^2}\,,
\end{equation}
so that in the comoving frame 
$\mathcal{L}_{\text{int}}=-\mu_B\,\bigl(n_B-n_{\bar B}\bigr)$. This expression 
explicitly shows that a non-vanishing time derivative of the Ricci scalar acts 
as a background field that differentiates between baryons and antibaryons, 
inducing opposite effective chemical potentials, $\mu_B = -\mu_{\bar B}$.

From a symmetry standpoint, it is worth noting that the baryon current \(J^0\) 
changes sign under CPT (\(J^0 \to -J^0\)), as does, in principle, the time 
derivative of the Ricci scalar (\(\dot{\mathcal{R}} \to -\dot{\mathcal{R}}\)). 
The interaction term \(\mathcal{L}_{\text{int}}\) is therefore formally CPT-even. 
However, in a cosmological background where \(\dot{\mathcal{R}}\) acquires a 
definite time-dependent value, it is treated as a fixed classical quantity that 
does not transform under CPT. In this case, the term 
\(J^\mu \partial_\mu \mathcal{R} \simeq J^0 \dot{\mathcal{R}}\) behaves 
effectively as a CPT-odd interaction (see Tab.~\ref{TabI}, which
summarises the transformation properties of the interaction~\eqref{Jmu} in the 
comoving
frame of the cosmic fluid within a spatially homogeneous FRW background, with
the curvature held fixed under the discrete symmetries).  
This results in an \emph{effective} CPT violation, manifested as an energy 
splitting between baryons and antibaryons. 
The corresponding dynamically induced chemical potential enables the generation 
of a net baryon asymmetry even in thermal equilibrium, as discussed before.

\begin{table}[t]
\centering
\begin{tabular}{c|c|c|c}
\hline
\textbf{Symmetry} & \textbf{$J^{0}$} & \textbf{$\dot{\mathcal R}\, 
\text{(fixed)}$} &
\textbf{$\mathcal{L}_{\text{int}}$} \\
\hline
C   & $-$ & +    & $-$ \\
P   & $+$ & +    & $+$ \\
T   & $+$ & +    & $+$ \\
CPT & $-$ & +    & $-$ \\
\hline
\end{tabular}
\caption{CPT transformation properties of the interaction~\eqref{Jmu} in the 
comoving
frame of the fluid within the FRW metric, with
the curvature held fixed under the  symmetries.}
\label{TabI}
\end{table}

Now, since baryogenesis is expected to take place shortly after reheating, the 
Universe is well described by a radiation-dominated
FRW background filled with an ultrarelativistic plasma. In this regime most 
species carrying baryon number are effectively massless and remain in thermal
equilibrium, so that the small chemical potential $\mu_B(t)$ can be treated as
a perturbation. Under these conditions one can use the standard equilibrium
expression for the net baryon number density,
\begin{equation}
\label{diff}
n_{B}-n_{\bar B}=\frac{g_{B}}{6}\,\mu_{B}\,T^{2}\,,
\end{equation}
where $g_{B}\sim\mathcal{O}(1)$ denotes the number of effectively relativistic 
degrees of freedom
carrying baryon number.

Following the standard convention in the baryogenesis 
literature~\cite{Davoudiasl:2013pda,Lambiase:2006dq,Fukushima:2016wyz,
Odintsov:2016hgc,Ramos:2017cot,Luciano:2022pzg,Luciano:2022ely,Das:2021nbq,
Troisi:2025ksj}, we now characterise the generated
asymmetry by the normalised quantity
\begin{equation}
    \eta_B \equiv\left.\frac{n_B - n_{\bar B}}{s}\right|_{T=T_D}\,,
\end{equation}
where $T_D$ is the
decoupling (or freeze-out) temperature of the baryon-number-violating
interactions and
$s$ denotes the entropy density of the plasma. 
it is given by $s = \dfrac{2\pi^2}{45}\,g_*\,T^3$, with $g_*\simeq106$ the 
effective number of relativistic degrees of freedom. 

Two comments are in order. First, the asymmetry should be evaluated at 
$T=T_{D}$, since above this temperature the baryon-number-violating 
interactions are sufficiently rapid to maintain chemical equilibrium and drive 
$n_{B}-n_{\bar B}$ toward its equilibrium value, whereas below $T_{D}$ they 
become inefficient and the ratio $(n_{B}-n_{\bar B})/s$ is frozen in and 
remains approximately constant during the subsequent adiabatic expansion of 
the Universe.

Furthermore, regarding the entropy density, we use the standard 
expression for a relativistic plasma,
$s=\frac{2\pi^{2}g_{*}}{45}\,T^{3}$, 
where $g_{*}\simeq106$ denotes the effective number of relativistic degrees of 
freedom. 
We emphasise that, in our framework, the modification of the 
Bekenstein-Hawking entropy of the cosmological horizon does not affect the 
thermodynamic entropy density of the plasma. Indeed, as long as the microscopic 
particle content and its thermal equilibrium distribution remain standard, 
the expression above for $s$ remains valid and can be consistently used to 
normalise the baryon asymmetry.

By using Eq.~\eqref{diff} along with the definition~\eqref{pot}, the parameter 
$\eta_B$ takes the final form
\begin{equation}
\label{etbet}
    \eta_B=-\left.\frac{15g_B}{4\pi^2g_*}\,\frac{\dot 
R}{M_*^2\,T}\right|_{T=T_D}\,,
\end{equation}
which indicates that the baryon asymmetry parameter is different from zero 
provided that $\dot R \neq 0$.

In the framework of GR, the Ricci scalar \(R\) for a flat FRW Universe is given 
by 
\begin{equation}
\label{Ricci}
    R=6\left(\dot H+2H^2\right)=8\pi G\left(\rho-3p\right),
\end{equation}
where in the second step we have used the standard Friedmann 
equations~\eqref{F1} and~\eqref{F2}. During the radiation-dominated era, the 
cosmological fluid has an adiabatic index \(w = 1/3\) (i.e.\ \(p=\rho/3\)), 
which leads to a vanishing Ricci scalar, \(R=0\). Consequently, its time 
derivative also vanishes, \(\dot R=0\), and gravitational baryogenesis cannot 
produce a baryon asymmetry in this regime (\(\eta_B\propto\dot R=0\)).

This perspective fundamentally changes 
in cosmological scenarios that admit a non-vanishing time derivative of the 
Ricci scalar. This situation arises in the modified 
entropic model introduced in the following subsection.

\subsection{Generalzed MHR-driven baryogenesis}

Let us now examine the mechanism of gravitational baryogenesis within the 
framework of cosmology based on the generalized mass-to-horizon 
entropy~\eqref{GMHE}. We mention that, although in the present work 
we adopt a different entropic model, we do not modify the underlying Lagrangian 
of the gravitational theory. Consequently, the field equations formally retain 
their standard GR form, and the fundamental geometric quantities, such as the 
Ricci scalar, preserve their usual formal expressions. In other words, we assume 
that the entropic corrections act only as effective modifications of the 
cosmological dynamics, without introducing new metric degrees of freedom. This 
assumption is reasonably justified at first order, since we consider only small 
deviations of the generalized entropy from the standard Bekenstein-Hawking case, 
consistent with the available observational 
constraints~\cite{Basilakos:2025wwu,Luciano:2025ovj}. A more complete analysis, 
starting from a modified gravitational action and deriving in a systematic way 
the corresponding field equations and the new expressions for the geometric 
quantities will be developed as a natural extension of the present work.

Using the modified Friedmann equations~\eqref{MFr1} and~\eqref{MFr2}, the Ricci 
scalar~\eqref{Ricci} in the radiation-dominated era becomes
\begin{equation}    
R(t)=\frac{2^{\frac{10-2n}{3-n}}\left(1-n\right)}{{\left(1-\frac{n}{3}\right)}^{
\frac{1-n}{3-n}}}\left[\frac{\pi 
G\hspace{0.2mm}\rho(t)}{n\gamma}\right]^{\frac{2}{3-n}}\,,
\end{equation}
which is generally non-vanishing for $n\neq1$.
In turn, its time derivative reads
\begin{equation}
    \dot R=\frac{2^{\frac{21-5n}{3-n}}\times 
3^{\frac{n}{n-3}}\left(n-1\right)}{\left(3-n\right)^{\frac{3-2n}{3-n}}}
\left(\frac{\pi G\hspace{0.2mm}\rho}{n\gamma}\right)^{\frac{3}{3-n}}\,,
    \label{derivative}
\end{equation}
where we have used the continuity equation~\eqref{cont} together with the 
radiation equation of state $p=\rho/3$.

Therefore, even in a radiation background the modified entropic framework leads 
to a nontrivial, dynamically evolving curvature scalar, which can serve as the 
source required for gravitational baryogenesis~\eqref{etbet}.

In order to compute relation \eqref{etbet} in light of \eqref{derivative}, 
the matter energy density~$\rho$ must be specified. Since our analysis focuses 
on the radiation-dominated era, it is thermodynamically consistent to identify 
the energy density of the effective fluid with that of the relativistic 
particle species, as provided by the Boltzmann equation~\cite{Kolb:1983ni}
\begin{equation}
\label{rho}
    \rho=\frac{\pi^2 g_*}{30} T^4\,.
\end{equation}
By substituting Eqs.~\eqref{derivative} and \eqref{rho} into~\eqref{etbet}, we 
obtain
\begin{equation}
\label{oureta}
    \eta_B=c_n\,g_B\,g_*^{\frac{n}{3-n}} 
M_p^{\frac{2\left(n-6\right)}{3-n}}T_D^{\frac{9+n}{3-n}}\,,
\end{equation}
where we have defined 
\begin{equation}
c_{n\,\gamma}\equiv2^{\frac{21-6n}{3-n}}\times45^{\frac{n}{n-3}}\times\pi^{\frac{6+n}{3-n}
}\left(3-n\right)^{\frac{2n-3}{3-n}}\left(1-n\right)\left(n\gamma\right)^{\frac{3}{n-3}}\,.
\end{equation}
Furthermore,  we have set $M_*=(8\pi G)^{-1/2}$, with $G=1/M_p^2$ in our units. 
It is straightforward to verify that, in the limit \(n=1\), the coefficient 
\(c_n\) vanishes, thus recovering the trivial GR behaviour of $\eta_B$ in the 
radiation-dominated era. 

For the purpose of comparing our results with observational bounds on the 
parameter~\(\eta_B\), we note that Eq.~\eqref{oureta} becomes more manageable 
when small deviations of \(n\) from unity are considered, as supported by 
observational evidence in Refs.~\cite{Basilakos:2025wwu,Luciano:2025ovj}. 
To this end, and in order to extract a constraint on the parameter~$n$, it is convenient to rewrite the decoupling temperature~$T_D$ 
as follows:
\begin{equation}
\frac{T_D}{M_p}= 10^{-3} x\,,
\end{equation}
where $M_p\simeq10^{19}\,\text{GeV}$. 
Thus, in order for the decoupling temperature to satisfy the condition 
$T_D \lesssim M_I = 3.3 \times 10^{16}\,\mathrm{GeV}$, where $M_I$ denotes the inflationary scale constrained by the upper bound on tensor mode fluctuations~\cite{Davoudiasl:2004gf}, 
we must set $0 < x \lesssim 3.3$.

Furthermore, to compare our bound on~$n$ with those recently obtained in the literature 
(see, e.g.,~\cite{Basilakos:2025wwu,Luciano:2025ovj,Luciano:2025tio}), 
we adopt the same unit conventions and work with~$\gamma = 1$ in units where 
$8\pi G = 1$, which implies $M_p=\sqrt{8\pi}$.
In doing so, Eq.~\eqref{oureta} takes the equivalent form
\begin{equation}
\label{neweta}
    \eta_B=\tilde c_n\, g_B\, g_*^{\frac{n}{3-n}}\,,
\end{equation}
with
\begin{eqnarray}
\nonumber
    \tilde c_n&\equiv &2^{\frac{3\left(7+3n\right)}{2\left(n-3\right)}}\times 5^{\frac{27+4n}{n-3}}\times 9^{\frac{n}{n-3}}\times \pi^{\frac{9+5n}{2\left(3-n\right)}}\times x^{\frac{9+n}{3-n}}
    \\[2mm]
    &&\times\left(3-n\right)^{\frac{2n-3}{3-n}}\left(1-n\right)n^{\frac{3}{n-3}}\,.
\end{eqnarray}
At this point, we can expand Eq.~\eqref{neweta} to leading order in \((n-1)\) to obtain
\begin{equation}
\label{expan}
    \eta_B\simeq 1.05\times 10^{-12} g_B\,\sqrt{g_*}\left(1-n\right)x^5
    \,+\, \mathcal{O}\left(n-1\right)^2\,.
\end{equation}
It is therefore evident that the modified entropy~\eqref{GMHE} can generate a 
non-zero baryon asymmetry as a result of the modifications it introduces to the 
Friedmann equations.

\begin{figure}[t]
\centering\includegraphics[width=0.48\textwidth]{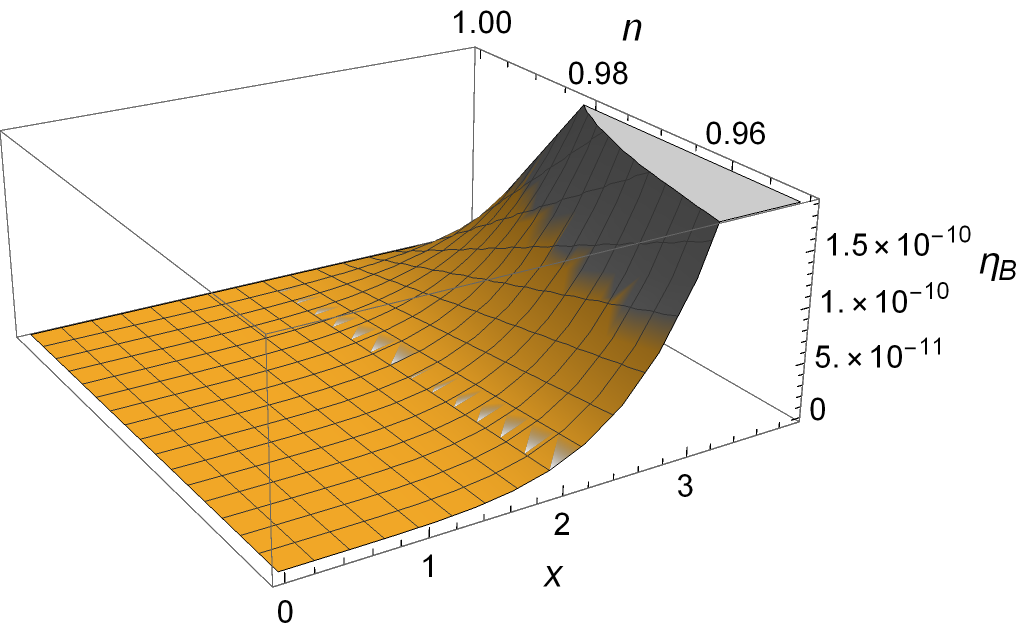}\caption{3D plot of 
$\eta_B$ in Eq.~\eqref{expan}. The grey 
region is excluded by the observational bound \( \eta_B \lesssim 10^{-10} 
\)~\cite{ParticleDataGroup:2018ovx}.}%
\label{Fig1}%
\end{figure}

The 3D plot of $\eta_B$ is displayed in Fig.~\ref{Fig1} as a function of 
$n$ and $x$. The grey region marks the 
portion of parameter space excluded by the observational bound \( \eta_B 
\lesssim \eta_{\rm obs} \sim 10^{-10} \)~\cite{ParticleDataGroup:2018ovx}, which 
is derived from the baryon-to-photon ratio inferred from CMB anisotropies and 
primordial light-element abundances from Big Bang Nucleosynthesis.

To enable a more quantitative analysis, we consider a benchmark scenario in which the decoupling temperature is fixed at its conventional upper limit, 
\(T_{D} = M_{I}\)~\cite{Das:2021nbq,Luciano:2022ely,Luciano:2022pzg,Troisi:2025ksj}, which corresponds to \(x = 3.3\). Within this framework, the observational constraint on \(\eta_{B}\) can be 
consistently translated into a quantitative exclusion bound on the parameter 
\(n\), thereby delineating the phenomenologically viable region of the parameter 
space. By substituting this value of $x$ into Eq.~\eqref{expan} and setting 
$g_*\simeq106, g_B\sim\mathcal{O}(1)$, the 
condition $0<\eta_B\lesssim10^{-10}$ implies 
\begin{equation}
\label{result}
    0.98\lesssim n<1\,,
\end{equation}
which sets the limit $0<1-n \lesssim \mathcal{O}(10^{-2})$ on the deviation from 
the area-law scaling of the horizon entropy. Based on the discussion below 
Eq.~\eqref{GMHE}, this result can be interpreted as indicating that the entropy 
associated with the apparent horizon grows sub-extensively ($n<1$) with its 
area. In this regime, the number of microscopic gravitational degrees of 
freedom contributing to the entropy increases more slowly than the horizon area 
itself.  

Such a sub-extensive scaling represents a mild departure from the standard 
holographic extensivity implied by the Bekenstein-Hawking relation and may 
signal the onset of nonlocal or quantum-gravitational corrections to the 
semiclassical description of spacetime thermodynamics. The bound $1-n \lesssim 
\mathcal{O}(10^{-2})$ then quantifies how close the system remains to 
holographic extensivity, while still allowing for a small, dynamically relevant 
modification of the entropy-area relation at cosmological scales.

\subsection{Comparison with previous constraints}

We close this section with a  comparison with other recent bounds on the 
entropic exponent $n$ (see Tab.~\ref{TabII} for a summary). In order to acheive 
that  we recall that the observational confrontation of the modified 
cosmological equations~\eqref{FM1},\eqref{FM2} with Type~Ia Supernovae 
(SNe~Ia), Cosmic Chronometers (CC), Baryon Acoustic Oscillations (BAO) 
(including the DESI DR2 release), and the Supernovae $H_0$ for the Equation of 
State (SH0ES) data yields the best-fit value 
$n = 0.945^{+0.070}_{-0.070}$,
in good agreement with the result obtained in the present 
analysis~\cite{Luciano:2025ovj}. 
This constraint relies exclusively on 
late-time cosmological probes, thus neglecting the early-universe physics 
encoded in the Cosmic Microwave Background (CMB). 
As such, it primarily tests the geometric sector of the entropic cosmology, 
where the modifications to the horizon entropy affect the background expansion 
but do not alter the acoustic physics of the primordial plasma.

Furthermore, our constraint slightly improves upon the bound $n > 
0.884^{+0.002}_{-0.001}$, obtained from the analysis of the primordial 
gravitational wave (PGW) spectrum constrained by Pulsar Timing Array (PTA) 
observations~\cite{Luciano:2025tio}. 
This result probes the entropic dynamics at much earlier epochs, revealing that 
deviations from the area law may have been more pronounced in the 
pre-recombination Universe, where quantum-gravitational and radiation couplings 
are expected to dominate.

In this context, we note that a more comprehensive analysis of the same 
generalized entropic cosmology, allowing for a free scaling parameter $\gamma$, 
was presented in Ref.~\cite{Gohar:2023lta}. By performing a joint fit to SNe~Ia, 
CC, BAO, Gamma-Ray Burst (GRB) and CMB data, it was shown that the model is 
fully equivalent to the standard $\Lambda$CDM cosmology for $n = 3$, while for 
$\log \gamma < -3$, and irrespective of the value of $n$, the model displays 
excellent agreement with the observational data, yielding results that are 
statistically indistinguishable from $\Lambda$CDM in Bayesian terms.

\begin{table}[t]
\centering
\caption{Bounds on the entropic parameters.}
\resizebox{\columnwidth}{!}{%
\begin{tabular}{ccc}
\hline\hline
\textbf{\Large{Dataset}} & \boldmath{\Large{$\gamma$}} & \textbf{\Large{$n$}} 
\\[3pt]
\hline
\Large{Baryogenesis (this work)} & \Large{$1$} & \Large{$[0.98,1[$} \\[6pt]
\Large{SNIa+CC+BAO(DESI DR2)+SH0ES~\cite{Luciano:2025ovj}} & \Large{${1}$} & 
\Large{${0.945\pm0.070}$} \\[6pt]
\Large{PGWs~\cite{Luciano:2025tio}} & \Large{$1$} & \Large{$\gtrsim 
0.884^{+0.002}_{-0.001}$} \\[6pt]
\Large{CC+SNIa+BAO(DESI DR1)~\cite{Basilakos:2025wwu}} & \Large{$1$} & 
\Large{$1.09 \pm 0.01$} \\[6pt]
\Large{SNIa+CC+BAO+GRB+CMB~\cite{Gohar:2023lta}} & \Large{($< e^{-3}$)} & 
\Large{Any} \\[3pt]
\hline\hline
\end{tabular}%
}
\label{TabII}
\end{table}

\section{Discussion and Conclusions}
\label{Conc}
Cosmological observations indicate a persistent baryon asymmetry that cannot be 
satisfactorily explained within the framework of standard cosmology. This 
discrepancy strongly suggests the presence of physics beyond the conventional 
paradigm. In the present work we investigated how such an asymmetry can arise 
from modifications to the Friedmann equations induced by a generalized 
mass-to-horizon entropy, thereby providing a potential mechanism for 
baryogenesis within a non-standard entropic framework.

The generalized mass-to-horizon entropy arises from a modified relation between 
the mass enclosed by the cosmological horizon and its radius, constructed to 
preserve consistency with the Clausius relation. This formulation introduces a 
power-law deviation from the standard Bekenstein-Hawking area law, quantified by 
the entropic index $n$. By applying the gravity-thermodynamics conjecture, the 
ensuing corrections to the Friedmann equations alter the evolution of the Hubble 
parameter during the radiation-dominated epoch. As a result, even the standard 
supergravity coupling between the Ricci scalar and the baryon current can 
generate a non-vanishing baryon asymmetry in the early Universe.

As we have shown, generating the observed baryon asymmetry within the framework 
of the generalized mass-to-horizon-entropy cosmology requires the entropic 
exponent to lie within the range $0.98 \lesssim n < 1$, corresponding to a 
sub-extensive deviation from the standard Bekenstein-Hawking entropy. This 
constraint improves upon the bounds previously obtained in the literature, while 
confirming that only small departures from holographic extensivity are 
compatible with current observations.

It is worth noting that an independent analysis based on CC + SNIa + BAO 
measurements has recently reported a best-fit value of $n>1$, corresponding to a 
superextensive scaling of the horizon entropy  \cite{Basilakos:2025wwu} (see 
also 
Tab.~\ref{TabII}). 
Understanding how this result may be reconciled with the sub-extensive regime 
found in the present baryogenesis framework is an interesting open 
question. One possible interpretation is that the entropic exponent $n$ does not 
correspond to a fixed universal constant, but rather to an effective, 
scale-dependent quantity that evolves with the thermodynamic state of the 
Universe. In this speculative picture, the dynamical evolution of $n$ would 
reflect the changing influence of microscopic gravitational degrees of freedom 
as cosmic expansion progresses. A thorough theoretical investigation and joint 
analysis of early- and late-time data will be required to determine whether this 
interpretation can coherently unify current observational bounds within a 
consistent entropic-cosmological framework. 
Work in this direction is currently in progress and will be developed elsewhere.

\acknowledgments 
The research of GGL is supported by the postdoctoral fellowship program of the 
University of Lleida. GGL and ENS gratefully acknowledge  the contribution of 
the LISA 
Cosmology Working Group (CosWG), as well as support from the COST Actions 
CA21136 - \textit{Addressing observational tensions in cosmology with 
systematics and fundamental physics (CosmoVerse)} - CA23130, \textit{Bridging 
high and low energies in search of quantum gravity (BridgeQG)} and CA21106 -  
\textit{COSMIC WISPers in the Dark Universe: Theory, astrophysics and 
experiments (CosmicWISPers)}.

\bibliography{Bib}

\begin{thebibliography}{96}
\expandafter\ifx\csname natexlab\endcsname\relax\def\natexlab#1{#1}\fi
\expandafter\ifx\csname bibnamefont\endcsname\relax
  \def\bibnamefont#1{#1}\fi
\expandafter\ifx\csname bibfnamefont\endcsname\relax
  \def\bibfnamefont#1{#1}\fi
\expandafter\ifx\csname citenamefont\endcsname\relax
  \def\citenamefont#1{#1}\fi
\expandafter\ifx\csname url\endcsname\relax
  \def\url#1{\texttt{#1}}\fi
\expandafter\ifx\csname urlprefix\endcsname\relax\def\urlprefix{URL }\fi
\providecommand{\bibinfo}[2]{#2}
\providecommand{\eprint}[2][]{\url{#2}}

\bibitem[{\citenamefont{Riess et~al.}(1998)}]{SupernovaSearchTeam:1998fmf}
\bibinfo{author}{\bibfnamefont{A.~G.} \bibnamefont{Riess}} \bibnamefont{et~al.} (\bibinfo{collaboration}{Supernova Search Team}), \bibinfo{journal}{Astron. J.} \textbf{\bibinfo{volume}{116}}, \bibinfo{pages}{1009} (\bibinfo{year}{1998}), \eprint{astro-ph/9805201}.

\bibitem[{\citenamefont{Perlmutter et~al.}(1999)}]{SupernovaCosmologyProject:1998vns}
\bibinfo{author}{\bibfnamefont{S.}~\bibnamefont{Perlmutter}} \bibnamefont{et~al.} (\bibinfo{collaboration}{Supernova Cosmology Project}), \bibinfo{journal}{Astrophys. J.} \textbf{\bibinfo{volume}{517}}, \bibinfo{pages}{565} (\bibinfo{year}{1999}), \eprint{astro-ph/9812133}.

\bibitem[{\citenamefont{Smoot et~al.}(1992)}]{COBE:1992syq}
\bibinfo{author}{\bibfnamefont{G.~F.} \bibnamefont{Smoot}} \bibnamefont{et~al.} (\bibinfo{collaboration}{COBE}), \bibinfo{journal}{Astrophys. J. Lett.} \textbf{\bibinfo{volume}{396}}, \bibinfo{pages}{L1} (\bibinfo{year}{1992}).

\bibitem[{\citenamefont{Bennett et~al.}(2003)}]{WMAP:2003ivt}
\bibinfo{author}{\bibfnamefont{C.~L.} \bibnamefont{Bennett}} \bibnamefont{et~al.} (\bibinfo{collaboration}{WMAP}), \bibinfo{journal}{Astrophys. J. Suppl.} \textbf{\bibinfo{volume}{148}}, \bibinfo{pages}{1} (\bibinfo{year}{2003}), \eprint{astro-ph/0302207}.

\bibitem[{\citenamefont{Colless et~al.}(2001)}]{2DFGRS:2001zay}
\bibinfo{author}{\bibfnamefont{M.}~\bibnamefont{Colless}} \bibnamefont{et~al.} (\bibinfo{collaboration}{2DFGRS}), \bibinfo{journal}{Mon. Not. Roy. Astron. Soc.} \textbf{\bibinfo{volume}{328}}, \bibinfo{pages}{1039} (\bibinfo{year}{2001}), \eprint{astro-ph/0106498}.

\bibitem[{\citenamefont{Tegmark et~al.}(2004)}]{SDSS:2003eyi}
\bibinfo{author}{\bibfnamefont{M.}~\bibnamefont{Tegmark}} \bibnamefont{et~al.} (\bibinfo{collaboration}{SDSS}), \bibinfo{journal}{Phys. Rev. D} \textbf{\bibinfo{volume}{69}}, \bibinfo{pages}{103501} (\bibinfo{year}{2004}), \eprint{astro-ph/0310723}.

\bibitem[{\citenamefont{Alam et~al.}(2017)}]{BOSS:2016wmc}
\bibinfo{author}{\bibfnamefont{S.}~\bibnamefont{Alam}} \bibnamefont{et~al.} (\bibinfo{collaboration}{BOSS}), \bibinfo{journal}{Mon. Not. Roy. Astron. Soc.} \textbf{\bibinfo{volume}{470}}, \bibinfo{pages}{2617} (\bibinfo{year}{2017}), \eprint{1607.03155}.

\bibitem[{\citenamefont{Capozziello and De~Laurentis}(2011)}]{Capozziello:2011et}
\bibinfo{author}{\bibfnamefont{S.}~\bibnamefont{Capozziello}} \bibnamefont{and} \bibinfo{author}{\bibfnamefont{M.}~\bibnamefont{De~Laurentis}}, \bibinfo{journal}{Phys. Rept.} \textbf{\bibinfo{volume}{509}}, \bibinfo{pages}{167} (\bibinfo{year}{2011}), \eprint{1108.6266}.

\bibitem[{\citenamefont{Saridakis et~al.}(2021)}]{CANTATA:2021asi}
\bibinfo{author}{\bibfnamefont{E.~N.} \bibnamefont{Saridakis}} \bibnamefont{et~al.} (\bibinfo{collaboration}{CANTATA}), \emph{\bibinfo{title}{{Modified Gravity and Cosmology. An Update by the CANTATA Network}}} (\bibinfo{publisher}{Springer}, \bibinfo{year}{2021}), ISBN \bibinfo{isbn}{978-3-030-83714-3, 978-3-030-83717-4, 978-3-030-83715-0}, \eprint{2105.12582}.

\bibitem[{\citenamefont{Olive}(1990)}]{Olive:1989nu}
\bibinfo{author}{\bibfnamefont{K.~A.} \bibnamefont{Olive}}, \bibinfo{journal}{Phys. Rept.} \textbf{\bibinfo{volume}{190}}, \bibinfo{pages}{307} (\bibinfo{year}{1990}).

\bibitem[{\citenamefont{Bartolo et~al.}(2004)\citenamefont{Bartolo, Komatsu, Matarrese, and Riotto}}]{Bartolo:2004if}
\bibinfo{author}{\bibfnamefont{N.}~\bibnamefont{Bartolo}}, \bibinfo{author}{\bibfnamefont{E.}~\bibnamefont{Komatsu}}, \bibinfo{author}{\bibfnamefont{S.}~\bibnamefont{Matarrese}}, \bibnamefont{and} \bibinfo{author}{\bibfnamefont{A.}~\bibnamefont{Riotto}}, \bibinfo{journal}{Phys. Rept.} \textbf{\bibinfo{volume}{402}}, \bibinfo{pages}{103} (\bibinfo{year}{2004}), \eprint{astro-ph/0406398}.

\bibitem[{\citenamefont{Copeland et~al.}(2006)\citenamefont{Copeland, Sami, and Tsujikawa}}]{Copeland:2006wr}
\bibinfo{author}{\bibfnamefont{E.~J.} \bibnamefont{Copeland}}, \bibinfo{author}{\bibfnamefont{M.}~\bibnamefont{Sami}}, \bibnamefont{and} \bibinfo{author}{\bibfnamefont{S.}~\bibnamefont{Tsujikawa}}, \bibinfo{journal}{Int. J. Mod. Phys. D} \textbf{\bibinfo{volume}{15}}, \bibinfo{pages}{1753} (\bibinfo{year}{2006}), \eprint{hep-th/0603057}.

\bibitem[{\citenamefont{Cai et~al.}(2010)\citenamefont{Cai, Saridakis, Setare, and Xia}}]{Cai:2009zp}
\bibinfo{author}{\bibfnamefont{Y.-F.} \bibnamefont{Cai}}, \bibinfo{author}{\bibfnamefont{E.~N.} \bibnamefont{Saridakis}}, \bibinfo{author}{\bibfnamefont{M.~R.} \bibnamefont{Setare}}, \bibnamefont{and} \bibinfo{author}{\bibfnamefont{J.-Q.} \bibnamefont{Xia}}, \bibinfo{journal}{Phys. Rept.} \textbf{\bibinfo{volume}{493}}, \bibinfo{pages}{1} (\bibinfo{year}{2010}), \eprint{0909.2776}.

\bibitem[{\citenamefont{Di~Valentino et~al.}(2025)}]{CosmoVerse:2025txj}
\bibinfo{author}{\bibfnamefont{E.}~\bibnamefont{Di~Valentino}} \bibnamefont{et~al.} (\bibinfo{collaboration}{CosmoVerse}) (\bibinfo{year}{2025}), \eprint{2504.01669}.

\bibitem[{\citenamefont{Jacobson}(1995)}]{Jacobson:1995ab}
\bibinfo{author}{\bibfnamefont{T.}~\bibnamefont{Jacobson}}, \bibinfo{journal}{Phys. Rev. Lett.} \textbf{\bibinfo{volume}{75}}, \bibinfo{pages}{1260} (\bibinfo{year}{1995}), \eprint{gr-qc/9504004}.

\bibitem[{\citenamefont{Padmanabhan}(2005)}]{Padmanabhan:2003gd}
\bibinfo{author}{\bibfnamefont{T.}~\bibnamefont{Padmanabhan}}, \bibinfo{journal}{Phys. Rept.} \textbf{\bibinfo{volume}{406}}, \bibinfo{pages}{49} (\bibinfo{year}{2005}), \eprint{gr-qc/0311036}.

\bibitem[{\citenamefont{Padmanabhan}(2010)}]{Padmanabhan:2009vy}
\bibinfo{author}{\bibfnamefont{T.}~\bibnamefont{Padmanabhan}}, \bibinfo{journal}{Rept. Prog. Phys.} \textbf{\bibinfo{volume}{73}}, \bibinfo{pages}{046901} (\bibinfo{year}{2010}), \eprint{0911.5004}.

\bibitem[{\citenamefont{Frolov and Kofman}(2003)}]{Frolov:2002va}
\bibinfo{author}{\bibfnamefont{A.~V.} \bibnamefont{Frolov}} \bibnamefont{and} \bibinfo{author}{\bibfnamefont{L.}~\bibnamefont{Kofman}}, \bibinfo{journal}{JCAP} \textbf{\bibinfo{volume}{05}}, \bibinfo{pages}{009} (\bibinfo{year}{2003}), \eprint{hep-th/0212327}.

\bibitem[{\citenamefont{Cai and Kim}(2005)}]{Cai:2005ra}
\bibinfo{author}{\bibfnamefont{R.-G.} \bibnamefont{Cai}} \bibnamefont{and} \bibinfo{author}{\bibfnamefont{S.~P.} \bibnamefont{Kim}}, \bibinfo{journal}{JHEP} \textbf{\bibinfo{volume}{02}}, \bibinfo{pages}{050} (\bibinfo{year}{2005}), \eprint{hep-th/0501055}.

\bibitem[{\citenamefont{Akbar and Cai}(2007)}]{Akbar:2006kj}
\bibinfo{author}{\bibfnamefont{M.}~\bibnamefont{Akbar}} \bibnamefont{and} \bibinfo{author}{\bibfnamefont{R.-G.} \bibnamefont{Cai}}, \bibinfo{journal}{Phys. Rev. D} \textbf{\bibinfo{volume}{75}}, \bibinfo{pages}{084003} (\bibinfo{year}{2007}), \eprint{hep-th/0609128}.

\bibitem[{\citenamefont{Cai and Cao}(2007)}]{Cai:2006rs}
\bibinfo{author}{\bibfnamefont{R.-G.} \bibnamefont{Cai}} \bibnamefont{and} \bibinfo{author}{\bibfnamefont{L.-M.} \bibnamefont{Cao}}, \bibinfo{journal}{Phys. Rev. D} \textbf{\bibinfo{volume}{75}}, \bibinfo{pages}{064008} (\bibinfo{year}{2007}), \eprint{gr-qc/0611071}.

\bibitem[{\citenamefont{Paranjape et~al.}(2006)\citenamefont{Paranjape, Sarkar, and Padmanabhan}}]{Paranjape:2006ca}
\bibinfo{author}{\bibfnamefont{A.}~\bibnamefont{Paranjape}}, \bibinfo{author}{\bibfnamefont{S.}~\bibnamefont{Sarkar}}, \bibnamefont{and} \bibinfo{author}{\bibfnamefont{T.}~\bibnamefont{Padmanabhan}}, \bibinfo{journal}{Phys. Rev. D} \textbf{\bibinfo{volume}{74}}, \bibinfo{pages}{104015} (\bibinfo{year}{2006}), \eprint{hep-th/0607240}.

\bibitem[{\citenamefont{Akbar and Cai}(2006)}]{Akbar:2006er}
\bibinfo{author}{\bibfnamefont{M.}~\bibnamefont{Akbar}} \bibnamefont{and} \bibinfo{author}{\bibfnamefont{R.-G.} \bibnamefont{Cai}}, \bibinfo{journal}{Phys. Lett. B} \textbf{\bibinfo{volume}{635}}, \bibinfo{pages}{7} (\bibinfo{year}{2006}), \eprint{hep-th/0602156}.

\bibitem[{\citenamefont{Jamil et~al.}(2010)\citenamefont{Jamil, Saridakis, and Setare}}]{Jamil:2009eb}
\bibinfo{author}{\bibfnamefont{M.}~\bibnamefont{Jamil}}, \bibinfo{author}{\bibfnamefont{E.~N.} \bibnamefont{Saridakis}}, \bibnamefont{and} \bibinfo{author}{\bibfnamefont{M.~R.} \bibnamefont{Setare}}, \bibinfo{journal}{Phys. Rev. D} \textbf{\bibinfo{volume}{81}}, \bibinfo{pages}{023007} (\bibinfo{year}{2010}), \eprint{0910.0822}.

\bibitem[{\citenamefont{Cai and Ohta}(2010)}]{Cai:2009ph}
\bibinfo{author}{\bibfnamefont{R.-G.} \bibnamefont{Cai}} \bibnamefont{and} \bibinfo{author}{\bibfnamefont{N.}~\bibnamefont{Ohta}}, \bibinfo{journal}{Phys. Rev. D} \textbf{\bibinfo{volume}{81}}, \bibinfo{pages}{084061} (\bibinfo{year}{2010}), \eprint{0910.2307}.

\bibitem[{\citenamefont{Easson et~al.}(2011)\citenamefont{Easson, Frampton, and Smoot}}]{Easson:2010av}
\bibinfo{author}{\bibfnamefont{D.~A.} \bibnamefont{Easson}}, \bibinfo{author}{\bibfnamefont{P.~H.} \bibnamefont{Frampton}}, \bibnamefont{and} \bibinfo{author}{\bibfnamefont{G.~F.} \bibnamefont{Smoot}}, \bibinfo{journal}{Phys. Lett. B} \textbf{\bibinfo{volume}{696}}, \bibinfo{pages}{273} (\bibinfo{year}{2011}), \eprint{1002.4278}.

\bibitem[{\citenamefont{Renyi}(1961)}]{renyi1961entropy}
\bibinfo{author}{\bibfnamefont{A.}~\bibnamefont{Renyi}}, in \emph{\bibinfo{booktitle}{Proceedings of the Fourth Berkeley Symposium on Mathematical Statistics and Probability}} (\bibinfo{publisher}{University of California Press}, \bibinfo{year}{1961}), vol.~\bibinfo{volume}{1}, pp. \bibinfo{pages}{547--561}.

\bibitem[{\citenamefont{Tsallis}(1988)}]{Tsallis:1987eu}
\bibinfo{author}{\bibfnamefont{C.}~\bibnamefont{Tsallis}}, \bibinfo{journal}{J. Statist. Phys.} \textbf{\bibinfo{volume}{52}}, \bibinfo{pages}{479} (\bibinfo{year}{1988}).

\bibitem[{\citenamefont{Tsallis}(2009)}]{Tsallis:2009}
\bibinfo{author}{\bibfnamefont{C.}~\bibnamefont{Tsallis}}, \emph{\bibinfo{title}{Introduction to Non-Extensive Statistical Mechanics: Approaching a Complex World}} (\bibinfo{publisher}{Springer}, \bibinfo{address}{Berlin}, \bibinfo{year}{2009}).

\bibitem[{\citenamefont{Sharma and Mittal}(1975)}]{Sharma1975}
\bibinfo{author}{\bibfnamefont{B.~D.} \bibnamefont{Sharma}} \bibnamefont{and} \bibinfo{author}{\bibfnamefont{D.~P.} \bibnamefont{Mittal}}, \bibinfo{journal}{Journal of Mathematical Sciences} \textbf{\bibinfo{volume}{10}}, \bibinfo{pages}{28} (\bibinfo{year}{1975}).

\bibitem[{\citenamefont{Kaniadakis}(2001)}]{kaniadakis2001non}
\bibinfo{author}{\bibfnamefont{G.}~\bibnamefont{Kaniadakis}}, \bibinfo{journal}{Physica A: Statistical mechanics and its applications} \textbf{\bibinfo{volume}{296}}, \bibinfo{pages}{405} (\bibinfo{year}{2001}).

\bibitem[{\citenamefont{Kaniadakis}(2002)}]{Kaniadakis:2002zz}
\bibinfo{author}{\bibfnamefont{G.}~\bibnamefont{Kaniadakis}}, \bibinfo{journal}{Phys. Rev. E} \textbf{\bibinfo{volume}{66}}, \bibinfo{pages}{056125} (\bibinfo{year}{2002}), \eprint{cond-mat/0210467}.

\bibitem[{\citenamefont{Luciano}(2024)}]{Luciano:2024bco}
\bibinfo{author}{\bibfnamefont{G.~G.} \bibnamefont{Luciano}}, \bibinfo{journal}{Eur. Phys. J. B} \textbf{\bibinfo{volume}{97}}, \bibinfo{pages}{80} (\bibinfo{year}{2024}), \eprint{2406.11373}.

\bibitem[{\citenamefont{Barrow}(2020)}]{Barrow:2020tzx}
\bibinfo{author}{\bibfnamefont{J.~D.} \bibnamefont{Barrow}}, \bibinfo{journal}{Phys. Lett. B} \textbf{\bibinfo{volume}{808}}, \bibinfo{pages}{135643} (\bibinfo{year}{2020}), \eprint{2004.09444}.

\bibitem[{\citenamefont{Hanel and Thurner}(2011)}]{hanel2011comprehensive}
\bibinfo{author}{\bibfnamefont{R.}~\bibnamefont{Hanel}} \bibnamefont{and} \bibinfo{author}{\bibfnamefont{S.}~\bibnamefont{Thurner}}, \bibinfo{journal}{Europhysics Letters} \textbf{\bibinfo{volume}{93}}, \bibinfo{pages}{20006} (\bibinfo{year}{2011}).

\bibitem[{\citenamefont{Lymperis and Saridakis}(2018)}]{Lymperis:2018iuz}
\bibinfo{author}{\bibfnamefont{A.}~\bibnamefont{Lymperis}} \bibnamefont{and} \bibinfo{author}{\bibfnamefont{E.~N.} \bibnamefont{Saridakis}}, \bibinfo{journal}{Eur. Phys. J. C} \textbf{\bibinfo{volume}{78}}, \bibinfo{pages}{993} (\bibinfo{year}{2018}), \eprint{1806.04614}.

\bibitem[{\citenamefont{Saridakis}(2020)}]{Saridakis:2020lrg}
\bibinfo{author}{\bibfnamefont{E.~N.} \bibnamefont{Saridakis}}, \bibinfo{journal}{JCAP} \textbf{\bibinfo{volume}{07}}, \bibinfo{pages}{031} (\bibinfo{year}{2020}), \eprint{2006.01105}.

\bibitem[{\citenamefont{Nojiri et~al.}(2019)\citenamefont{Nojiri, Odintsov, and Saridakis}}]{Nojiri:2019skr}
\bibinfo{author}{\bibfnamefont{S.}~\bibnamefont{Nojiri}}, \bibinfo{author}{\bibfnamefont{S.~D.} \bibnamefont{Odintsov}}, \bibnamefont{and} \bibinfo{author}{\bibfnamefont{E.~N.} \bibnamefont{Saridakis}}, \bibinfo{journal}{Eur. Phys. J. C} \textbf{\bibinfo{volume}{79}}, \bibinfo{pages}{242} (\bibinfo{year}{2019}), \eprint{1903.03098}.

\bibitem[{\citenamefont{Hern\'andez-Almada et~al.}(2022)\citenamefont{Hern\'andez-Almada, Leon, Maga\~na, Garc\'\i{}a-Aspeitia, Motta, Saridakis, Yesmakhanova, and Millano}}]{Hernandez-Almada:2021rjs}
\bibinfo{author}{\bibfnamefont{A.}~\bibnamefont{Hern\'andez-Almada}}, \bibinfo{author}{\bibfnamefont{G.}~\bibnamefont{Leon}}, \bibinfo{author}{\bibfnamefont{J.}~\bibnamefont{Maga\~na}}, \bibinfo{author}{\bibfnamefont{M.~A.} \bibnamefont{Garc\'\i{}a-Aspeitia}}, \bibinfo{author}{\bibfnamefont{V.}~\bibnamefont{Motta}}, \bibinfo{author}{\bibfnamefont{E.~N.} \bibnamefont{Saridakis}}, \bibinfo{author}{\bibfnamefont{K.}~\bibnamefont{Yesmakhanova}}, \bibnamefont{and} \bibinfo{author}{\bibfnamefont{A.~D.} \bibnamefont{Millano}}, \bibinfo{journal}{Mon. Not. Roy. Astron. Soc.} \textbf{\bibinfo{volume}{512}}, \bibinfo{pages}{5122} (\bibinfo{year}{2022}), \eprint{2112.04615}.

\bibitem[{\citenamefont{Dheepika et~al.}(2024)\citenamefont{Dheepika, T., and Mathew}}]{Dheepika:2022sio}
\bibinfo{author}{\bibfnamefont{M.}~\bibnamefont{Dheepika}}, \bibinfo{author}{\bibfnamefont{H.~B.~V.} \bibnamefont{T.}}, \bibnamefont{and} \bibinfo{author}{\bibfnamefont{T.~K.} \bibnamefont{Mathew}}, \bibinfo{journal}{Phys. Scripta} \textbf{\bibinfo{volume}{99}}, \bibinfo{pages}{015014} (\bibinfo{year}{2024}), \eprint{2211.14039}.

\bibitem[{\citenamefont{Jizba et~al.}(2022)\citenamefont{Jizba, Lambiase, Luciano, and Petruzziello}}]{Jizba:2022icu}
\bibinfo{author}{\bibfnamefont{P.}~\bibnamefont{Jizba}}, \bibinfo{author}{\bibfnamefont{G.}~\bibnamefont{Lambiase}}, \bibinfo{author}{\bibfnamefont{G.~G.} \bibnamefont{Luciano}}, \bibnamefont{and} \bibinfo{author}{\bibfnamefont{L.}~\bibnamefont{Petruzziello}}, \bibinfo{journal}{Phys. Rev. D} \textbf{\bibinfo{volume}{105}}, \bibinfo{pages}{L121501} (\bibinfo{year}{2022}), \eprint{2201.07919}.

\bibitem[{\citenamefont{Lambiase et~al.}(2023)\citenamefont{Lambiase, Luciano, and Sheykhi}}]{Lambiase:2023ryq}
\bibinfo{author}{\bibfnamefont{G.}~\bibnamefont{Lambiase}}, \bibinfo{author}{\bibfnamefont{G.~G.} \bibnamefont{Luciano}}, \bibnamefont{and} \bibinfo{author}{\bibfnamefont{A.}~\bibnamefont{Sheykhi}}, \bibinfo{journal}{Eur. Phys. J. C} \textbf{\bibinfo{volume}{83}}, \bibinfo{pages}{936} (\bibinfo{year}{2023}), \eprint{2307.04027}.

\bibitem[{\citenamefont{Jizba et~al.}(2024)\citenamefont{Jizba, Lambiase, Luciano, and Mastrototaro}}]{Jizba:2024klq}
\bibinfo{author}{\bibfnamefont{P.}~\bibnamefont{Jizba}}, \bibinfo{author}{\bibfnamefont{G.}~\bibnamefont{Lambiase}}, \bibinfo{author}{\bibfnamefont{G.~G.} \bibnamefont{Luciano}}, \bibnamefont{and} \bibinfo{author}{\bibfnamefont{L.}~\bibnamefont{Mastrototaro}}, \bibinfo{journal}{Eur. Phys. J. C} \textbf{\bibinfo{volume}{84}}, \bibinfo{pages}{1076} (\bibinfo{year}{2024}), \eprint{2403.09797}.

\bibitem[{\citenamefont{Ebrahimi and Sheykhi}(2024)}]{Ebrahimi:2024zrk}
\bibinfo{author}{\bibfnamefont{E.}~\bibnamefont{Ebrahimi}} \bibnamefont{and} \bibinfo{author}{\bibfnamefont{A.}~\bibnamefont{Sheykhi}}, \bibinfo{journal}{Phys. Dark Univ.} \textbf{\bibinfo{volume}{45}}, \bibinfo{pages}{101518} (\bibinfo{year}{2024}), \eprint{2405.13096}.

\bibitem[{\citenamefont{Nojiri et~al.}(2025)\citenamefont{Nojiri, Odintsov, Paul, and SenGupta}}]{Nojiri:2025gkq}
\bibinfo{author}{\bibfnamefont{S.}~\bibnamefont{Nojiri}}, \bibinfo{author}{\bibfnamefont{S.~D.} \bibnamefont{Odintsov}}, \bibinfo{author}{\bibfnamefont{T.}~\bibnamefont{Paul}}, \bibnamefont{and} \bibinfo{author}{\bibfnamefont{S.}~\bibnamefont{SenGupta}} (\bibinfo{year}{2025}), \eprint{2503.19056}.

\bibitem[{\citenamefont{Nojiri et~al.}(2021)\citenamefont{Nojiri, Odintsov, and Faraoni}}]{Nojiri:2021czz}
\bibinfo{author}{\bibfnamefont{S.}~\bibnamefont{Nojiri}}, \bibinfo{author}{\bibfnamefont{S.~D.} \bibnamefont{Odintsov}}, \bibnamefont{and} \bibinfo{author}{\bibfnamefont{V.}~\bibnamefont{Faraoni}}, \bibinfo{journal}{Phys. Rev. D} \textbf{\bibinfo{volume}{104}}, \bibinfo{pages}{084030} (\bibinfo{year}{2021}), \eprint{2109.05315}.

\bibitem[{\citenamefont{Gohar and Salzano}(2024{\natexlab{a}})}]{Gohar:2023lta}
\bibinfo{author}{\bibfnamefont{H.}~\bibnamefont{Gohar}} \bibnamefont{and} \bibinfo{author}{\bibfnamefont{V.}~\bibnamefont{Salzano}}, \bibinfo{journal}{Phys. Rev. D} \textbf{\bibinfo{volume}{109}}, \bibinfo{pages}{084075} (\bibinfo{year}{2024}{\natexlab{a}}), \eprint{2307.06239}.

\bibitem[{\citenamefont{Nojiri et~al.}(2022)\citenamefont{Nojiri, Odintsov, and Faraoni}}]{Nojiri:2022sfd}
\bibinfo{author}{\bibfnamefont{S.}~\bibnamefont{Nojiri}}, \bibinfo{author}{\bibfnamefont{S.~D.} \bibnamefont{Odintsov}}, \bibnamefont{and} \bibinfo{author}{\bibfnamefont{V.}~\bibnamefont{Faraoni}}, \bibinfo{journal}{Int. J. Geom. Meth. Mod. Phys.} \textbf{\bibinfo{volume}{19}}, \bibinfo{pages}{2250210} (\bibinfo{year}{2022}), \eprint{2207.07905}.

\bibitem[{\citenamefont{Gohar and Salzano}(2024{\natexlab{b}})}]{Gohar:2023hnb}
\bibinfo{author}{\bibfnamefont{H.}~\bibnamefont{Gohar}} \bibnamefont{and} \bibinfo{author}{\bibfnamefont{V.}~\bibnamefont{Salzano}}, \bibinfo{journal}{Phys. Lett. B} \textbf{\bibinfo{volume}{855}}, \bibinfo{pages}{138781} (\bibinfo{year}{2024}{\natexlab{b}}), \eprint{2307.01768}.

\bibitem[{\citenamefont{Basilakos et~al.}(2012)\citenamefont{Basilakos, Polarski, and Sola}}]{Basilakos:2012ra}
\bibinfo{author}{\bibfnamefont{S.}~\bibnamefont{Basilakos}}, \bibinfo{author}{\bibfnamefont{D.}~\bibnamefont{Polarski}}, \bibnamefont{and} \bibinfo{author}{\bibfnamefont{J.}~\bibnamefont{Sola}}, \bibinfo{journal}{Phys. Rev. D} \textbf{\bibinfo{volume}{86}}, \bibinfo{pages}{043010} (\bibinfo{year}{2012}), \eprint{1204.4806}.

\bibitem[{\citenamefont{Basilakos and Sola}(2014)}]{Basilakos:2014tha}
\bibinfo{author}{\bibfnamefont{S.}~\bibnamefont{Basilakos}} \bibnamefont{and} \bibinfo{author}{\bibfnamefont{J.}~\bibnamefont{Sola}}, \bibinfo{journal}{Phys. Rev. D} \textbf{\bibinfo{volume}{90}}, \bibinfo{pages}{023008} (\bibinfo{year}{2014}), \eprint{1402.6594}.

\bibitem[{\citenamefont{Tsallis and Cirto}(2013)}]{Tsallis:2013}
\bibinfo{author}{\bibfnamefont{C.}~\bibnamefont{Tsallis}} \bibnamefont{and} \bibinfo{author}{\bibfnamefont{L.~J.~L.} \bibnamefont{Cirto}}, \bibinfo{journal}{Eur. Phys. J. C} \textbf{\bibinfo{volume}{73}} (\bibinfo{year}{2013}).

\bibitem[{\citenamefont{Basilakos et~al.}(2025)\citenamefont{Basilakos, Lymperis, Petronikolou, and Saridakis}}]{Basilakos:2025wwu}
\bibinfo{author}{\bibfnamefont{S.}~\bibnamefont{Basilakos}}, \bibinfo{author}{\bibfnamefont{A.}~\bibnamefont{Lymperis}}, \bibinfo{author}{\bibfnamefont{M.}~\bibnamefont{Petronikolou}}, \bibnamefont{and} \bibinfo{author}{\bibfnamefont{E.~N.} \bibnamefont{Saridakis}} (\bibinfo{year}{2025}), \eprint{2503.24355}.

\bibitem[{\citenamefont{Luciano and Paliathanasis}(2025)}]{Luciano:2025ovj}
\bibinfo{author}{\bibfnamefont{G.~G.} \bibnamefont{Luciano}} \bibnamefont{and} \bibinfo{author}{\bibfnamefont{A.}~\bibnamefont{Paliathanasis}}, \bibinfo{journal}{Phys. Lett. B} \textbf{\bibinfo{volume}{870}}, \bibinfo{pages}{139954} (\bibinfo{year}{2025}), \eprint{2508.13260}.

\bibitem[{\citenamefont{Ormondroyd et~al.}(2025)\citenamefont{Ormondroyd, Handley, Hobson, and Lasenby}}]{Ormondroyd:2025iaf}
\bibinfo{author}{\bibfnamefont{A.~N.} \bibnamefont{Ormondroyd}}, \bibinfo{author}{\bibfnamefont{W.~J.} \bibnamefont{Handley}}, \bibinfo{author}{\bibfnamefont{M.~P.} \bibnamefont{Hobson}}, \bibnamefont{and} \bibinfo{author}{\bibfnamefont{A.~N.} \bibnamefont{Lasenby}} (\bibinfo{year}{2025}), \eprint{2503.17342}.

\bibitem[{\citenamefont{You et~al.}(2025)\citenamefont{You, Wang, and Yang}}]{You:2025uon}
\bibinfo{author}{\bibfnamefont{C.}~\bibnamefont{You}}, \bibinfo{author}{\bibfnamefont{D.}~\bibnamefont{Wang}}, \bibnamefont{and} \bibinfo{author}{\bibfnamefont{T.}~\bibnamefont{Yang}} (\bibinfo{year}{2025}), \eprint{2504.00985}.

\bibitem[{\citenamefont{Gu et~al.}(2025)}]{Gu:2025xie}
\bibinfo{author}{\bibfnamefont{G.}~\bibnamefont{Gu}} \bibnamefont{et~al.} (\bibinfo{year}{2025}), \eprint{2504.06118}.

\bibitem[{\citenamefont{Santos et~al.}(2025)\citenamefont{Santos, Morais, Pan, Yang, and Di~Valentino}}]{Santos:2025wiv}
\bibinfo{author}{\bibfnamefont{F.~B. M.~d.} \bibnamefont{Santos}}, \bibinfo{author}{\bibfnamefont{J.}~\bibnamefont{Morais}}, \bibinfo{author}{\bibfnamefont{S.}~\bibnamefont{Pan}}, \bibinfo{author}{\bibfnamefont{W.}~\bibnamefont{Yang}}, \bibnamefont{and} \bibinfo{author}{\bibfnamefont{E.}~\bibnamefont{Di~Valentino}} (\bibinfo{year}{2025}), \eprint{2504.04646}.

\bibitem[{\citenamefont{Li et~al.}(2025)\citenamefont{Li, Wang, Zhang, Saridakis, and Cai}}]{Li:2025cxn}
\bibinfo{author}{\bibfnamefont{C.}~\bibnamefont{Li}}, \bibinfo{author}{\bibfnamefont{J.}~\bibnamefont{Wang}}, \bibinfo{author}{\bibfnamefont{D.}~\bibnamefont{Zhang}}, \bibinfo{author}{\bibfnamefont{E.~N.} \bibnamefont{Saridakis}}, \bibnamefont{and} \bibinfo{author}{\bibfnamefont{Y.-F.} \bibnamefont{Cai}} (\bibinfo{year}{2025}), \eprint{2504.07791}.

\bibitem[{\citenamefont{Carloni et~al.}(2025)\citenamefont{Carloni, Luongo, and Muccino}}]{Carloni:2024zpl}
\bibinfo{author}{\bibfnamefont{Y.}~\bibnamefont{Carloni}}, \bibinfo{author}{\bibfnamefont{O.}~\bibnamefont{Luongo}}, \bibnamefont{and} \bibinfo{author}{\bibfnamefont{M.}~\bibnamefont{Muccino}}, \bibinfo{journal}{Phys. Rev. D} \textbf{\bibinfo{volume}{111}}, \bibinfo{pages}{023512} (\bibinfo{year}{2025}), \eprint{2404.12068}.

\bibitem[{\citenamefont{Luciano et~al.}(2026)\citenamefont{Luciano, Paliathanasis, and Saridakis}}]{Luciano:2025elo}
\bibinfo{author}{\bibfnamefont{G.~G.} \bibnamefont{Luciano}}, \bibinfo{author}{\bibfnamefont{A.}~\bibnamefont{Paliathanasis}}, \bibnamefont{and} \bibinfo{author}{\bibfnamefont{E.~N.} \bibnamefont{Saridakis}}, \bibinfo{journal}{JHEAp} \textbf{\bibinfo{volume}{49}}, \bibinfo{pages}{100427} (\bibinfo{year}{2026}), \eprint{2506.03019}.

\bibitem[{\citenamefont{Luciano et~al.}(2025)\citenamefont{Luciano, Paliathanasis, and Saridakis}}]{Luciano:2025hjn}
\bibinfo{author}{\bibfnamefont{G.~G.} \bibnamefont{Luciano}}, \bibinfo{author}{\bibfnamefont{A.}~\bibnamefont{Paliathanasis}}, \bibnamefont{and} \bibinfo{author}{\bibfnamefont{E.~N.} \bibnamefont{Saridakis}}, \bibinfo{journal}{JCAP} \textbf{\bibinfo{volume}{09}}, \bibinfo{pages}{013} (\bibinfo{year}{2025}), \eprint{2504.12205}.

\bibitem[{\citenamefont{Chaussidon et~al.}(2025)}]{Chaussidon:2025npr}
\bibinfo{author}{\bibfnamefont{E.}~\bibnamefont{Chaussidon}} \bibnamefont{et~al.} (\bibinfo{year}{2025}), \eprint{2503.24343}.

\bibitem[{\citenamefont{Anchordoqui et~al.}(2025)\citenamefont{Anchordoqui, Antoniadis, and Lust}}]{Anchordoqui:2025fgz}
\bibinfo{author}{\bibfnamefont{L.~A.} \bibnamefont{Anchordoqui}}, \bibinfo{author}{\bibfnamefont{I.}~\bibnamefont{Antoniadis}}, \bibnamefont{and} \bibinfo{author}{\bibfnamefont{D.}~\bibnamefont{Lust}} (\bibinfo{year}{2025}), \eprint{2503.19428}.

\bibitem[{\citenamefont{Ye and Cai}(2025)}]{Ye:2025ulq}
\bibinfo{author}{\bibfnamefont{G.}~\bibnamefont{Ye}} \bibnamefont{and} \bibinfo{author}{\bibfnamefont{Y.}~\bibnamefont{Cai}} (\bibinfo{year}{2025}), \eprint{2503.22515}.

\bibitem[{\citenamefont{Paliathanasis}(2025{\natexlab{a}})}]{Paliathanasis:2025dcr}
\bibinfo{author}{\bibfnamefont{A.}~\bibnamefont{Paliathanasis}}, \bibinfo{journal}{JCAP} \textbf{\bibinfo{volume}{09}}, \bibinfo{pages}{067} (\bibinfo{year}{2025}{\natexlab{a}}), \eprint{2503.20896}.

\bibitem[{\citenamefont{Yang et~al.}(2025)\citenamefont{Yang, Wang, Ren, Saridakis, and Cai}}]{Yang:2025mws}
\bibinfo{author}{\bibfnamefont{Y.}~\bibnamefont{Yang}}, \bibinfo{author}{\bibfnamefont{Q.}~\bibnamefont{Wang}}, \bibinfo{author}{\bibfnamefont{X.}~\bibnamefont{Ren}}, \bibinfo{author}{\bibfnamefont{E.~N.} \bibnamefont{Saridakis}}, \bibnamefont{and} \bibinfo{author}{\bibfnamefont{Y.-F.} \bibnamefont{Cai}} (\bibinfo{year}{2025}), \eprint{2504.06784}.

\bibitem[{\citenamefont{Paliathanasis}(2025{\natexlab{b}})}]{Paliathanasis:2025hjw}
\bibinfo{author}{\bibfnamefont{A.}~\bibnamefont{Paliathanasis}}, \bibinfo{journal}{Phys. Dark Univ.} \textbf{\bibinfo{volume}{49}}, \bibinfo{pages}{101993} (\bibinfo{year}{2025}{\natexlab{b}}), \eprint{2504.11132}.

\bibitem[{\citenamefont{Tyagi et~al.}(2025)\citenamefont{Tyagi, Haridasu, and Basak}}]{Tyagi:2025zov}
\bibinfo{author}{\bibfnamefont{U.~K.} \bibnamefont{Tyagi}}, \bibinfo{author}{\bibfnamefont{S.}~\bibnamefont{Haridasu}}, \bibnamefont{and} \bibinfo{author}{\bibfnamefont{S.}~\bibnamefont{Basak}} (\bibinfo{year}{2025}), \eprint{2504.11308}.

\bibitem[{\citenamefont{Kolb and Turner}(1983)}]{Kolb:1983ni}
\bibinfo{author}{\bibfnamefont{E.~W.} \bibnamefont{Kolb}} \bibnamefont{and} \bibinfo{author}{\bibfnamefont{M.~S.} \bibnamefont{Turner}}, \bibinfo{journal}{Ann. Rev. Nucl. Part. Sci.} \textbf{\bibinfo{volume}{33}}, \bibinfo{pages}{645} (\bibinfo{year}{1983}).

\bibitem[{\citenamefont{Shaposhnikov}(1987)}]{Shaposhnikov:1987tw}
\bibinfo{author}{\bibfnamefont{M.~E.} \bibnamefont{Shaposhnikov}}, \bibinfo{journal}{Nucl. Phys. B} \textbf{\bibinfo{volume}{287}}, \bibinfo{pages}{757} (\bibinfo{year}{1987}).

\bibitem[{\citenamefont{Sakharov}(1967)}]{Sakharov:1967dj}
\bibinfo{author}{\bibfnamefont{A.~D.} \bibnamefont{Sakharov}}, \bibinfo{journal}{Pisma Zh. Eksp. Teor. Fiz.} \textbf{\bibinfo{volume}{5}}, \bibinfo{pages}{32} (\bibinfo{year}{1967}).

\bibitem[{\citenamefont{Dolgov}(1992)}]{Dolgov:1991fr}
\bibinfo{author}{\bibfnamefont{A.~D.} \bibnamefont{Dolgov}}, \bibinfo{journal}{Phys. Rept.} \textbf{\bibinfo{volume}{222}}, \bibinfo{pages}{309} (\bibinfo{year}{1992}).

\bibitem[{\citenamefont{Cohen and Kaplan}(1987)}]{Cohen:1987vi}
\bibinfo{author}{\bibfnamefont{A.~G.} \bibnamefont{Cohen}} \bibnamefont{and} \bibinfo{author}{\bibfnamefont{D.~B.} \bibnamefont{Kaplan}}, \bibinfo{journal}{Phys. Lett. B} \textbf{\bibinfo{volume}{199}}, \bibinfo{pages}{251} (\bibinfo{year}{1987}).

\bibitem[{\citenamefont{Davoudiasl et~al.}(2004)\citenamefont{Davoudiasl, Kitano, Kribs, Murayama, and Steinhardt}}]{Davoudiasl:2004gf}
\bibinfo{author}{\bibfnamefont{H.}~\bibnamefont{Davoudiasl}}, \bibinfo{author}{\bibfnamefont{R.}~\bibnamefont{Kitano}}, \bibinfo{author}{\bibfnamefont{G.~D.} \bibnamefont{Kribs}}, \bibinfo{author}{\bibfnamefont{H.}~\bibnamefont{Murayama}}, \bibnamefont{and} \bibinfo{author}{\bibfnamefont{P.~J.} \bibnamefont{Steinhardt}}, \bibinfo{journal}{Phys. Rev. Lett.} \textbf{\bibinfo{volume}{93}}, \bibinfo{pages}{201301} (\bibinfo{year}{2004}), \eprint{hep-ph/0403019}.

\bibitem[{\citenamefont{Davoudiasl}(2013)}]{Davoudiasl:2013pda}
\bibinfo{author}{\bibfnamefont{H.}~\bibnamefont{Davoudiasl}}, \bibinfo{journal}{Phys. Rev. D} \textbf{\bibinfo{volume}{88}}, \bibinfo{pages}{095004} (\bibinfo{year}{2013}), \eprint{1308.3473}.

\bibitem[{\citenamefont{Lambiase and Scarpetta}(2006)}]{Lambiase:2006dq}
\bibinfo{author}{\bibfnamefont{G.}~\bibnamefont{Lambiase}} \bibnamefont{and} \bibinfo{author}{\bibfnamefont{G.}~\bibnamefont{Scarpetta}}, \bibinfo{journal}{Phys. Rev. D} \textbf{\bibinfo{volume}{74}}, \bibinfo{pages}{087504} (\bibinfo{year}{2006}), \eprint{astro-ph/0610367}.

\bibitem[{\citenamefont{Fukushima et~al.}(2016)\citenamefont{Fukushima, Mizuno, and Maeda}}]{Fukushima:2016wyz}
\bibinfo{author}{\bibfnamefont{M.}~\bibnamefont{Fukushima}}, \bibinfo{author}{\bibfnamefont{S.}~\bibnamefont{Mizuno}}, \bibnamefont{and} \bibinfo{author}{\bibfnamefont{K.-i.} \bibnamefont{Maeda}}, \bibinfo{journal}{Phys. Rev. D} \textbf{\bibinfo{volume}{93}}, \bibinfo{pages}{103513} (\bibinfo{year}{2016}), \eprint{1603.02403}.

\bibitem[{\citenamefont{Oikonomou and Saridakis}(2016)}]{Oikonomou:2016jjh}
\bibinfo{author}{\bibfnamefont{V.~K.} \bibnamefont{Oikonomou}} \bibnamefont{and} \bibinfo{author}{\bibfnamefont{E.~N.} \bibnamefont{Saridakis}}, \bibinfo{journal}{Phys. Rev. D} \textbf{\bibinfo{volume}{94}}, \bibinfo{pages}{124005} (\bibinfo{year}{2016}), \eprint{1607.08561}.

\bibitem[{\citenamefont{Odintsov and Oikonomou}(2016)}]{Odintsov:2016hgc}
\bibinfo{author}{\bibfnamefont{S.~D.} \bibnamefont{Odintsov}} \bibnamefont{and} \bibinfo{author}{\bibfnamefont{V.~K.} \bibnamefont{Oikonomou}}, \bibinfo{journal}{Phys. Lett. B} \textbf{\bibinfo{volume}{760}}, \bibinfo{pages}{259} (\bibinfo{year}{2016}), \eprint{1607.00545}.

\bibitem[{\citenamefont{Ramos and P{\'a}ramos}(2017)}]{Ramos:2017cot}
\bibinfo{author}{\bibfnamefont{M.~P. L.~P.} \bibnamefont{Ramos}} \bibnamefont{and} \bibinfo{author}{\bibfnamefont{J.}~\bibnamefont{P{\'a}ramos}}, \bibinfo{journal}{Phys. Rev. D} \textbf{\bibinfo{volume}{96}}, \bibinfo{pages}{104024} (\bibinfo{year}{2017}), \eprint{1709.04442}.

\bibitem[{\citenamefont{Luciano and Saridakis}(2022)}]{Luciano:2022pzg}
\bibinfo{author}{\bibfnamefont{G.~G.} \bibnamefont{Luciano}} \bibnamefont{and} \bibinfo{author}{\bibfnamefont{E.~N.} \bibnamefont{Saridakis}}, \bibinfo{journal}{Eur. Phys. J. C} \textbf{\bibinfo{volume}{82}}, \bibinfo{pages}{558} (\bibinfo{year}{2022}), \eprint{2203.12010}.

\bibitem[{\citenamefont{Luciano and Gine}(2022)}]{Luciano:2022ely}
\bibinfo{author}{\bibfnamefont{G.~G.} \bibnamefont{Luciano}} \bibnamefont{and} \bibinfo{author}{\bibfnamefont{J.}~\bibnamefont{Gine}}, \bibinfo{journal}{Phys. Lett. B} \textbf{\bibinfo{volume}{833}}, \bibinfo{pages}{137352} (\bibinfo{year}{2022}), \eprint{2204.02723}.

\bibitem[{\citenamefont{Das et~al.}(2022)\citenamefont{Das, Fridman, Lambiase, and Vagenas}}]{Das:2021nbq}
\bibinfo{author}{\bibfnamefont{S.}~\bibnamefont{Das}}, \bibinfo{author}{\bibfnamefont{M.}~\bibnamefont{Fridman}}, \bibinfo{author}{\bibfnamefont{G.}~\bibnamefont{Lambiase}}, \bibnamefont{and} \bibinfo{author}{\bibfnamefont{E.~C.} \bibnamefont{Vagenas}}, \bibinfo{journal}{Phys. Lett. B} \textbf{\bibinfo{volume}{824}}, \bibinfo{pages}{136841} (\bibinfo{year}{2022}), \eprint{2107.02077}.

\bibitem[{\citenamefont{Troisi et~al.}(2025)\citenamefont{Troisi, Lambiase, and Carloni}}]{Troisi:2025ksj}
\bibinfo{author}{\bibfnamefont{A.}~\bibnamefont{Troisi}}, \bibinfo{author}{\bibfnamefont{G.}~\bibnamefont{Lambiase}}, \bibnamefont{and} \bibinfo{author}{\bibfnamefont{S.}~\bibnamefont{Carloni}}, \bibinfo{journal}{JCAP} \textbf{\bibinfo{volume}{09}}, \bibinfo{pages}{043} (\bibinfo{year}{2025}), \eprint{2504.16751}.

\bibitem[{\citenamefont{Cai et~al.}(2009)\citenamefont{Cai, Cao, Hu, and Ohta}}]{Cai:2009qf}
\bibinfo{author}{\bibfnamefont{R.-G.} \bibnamefont{Cai}}, \bibinfo{author}{\bibfnamefont{L.-M.} \bibnamefont{Cao}}, \bibinfo{author}{\bibfnamefont{Y.-P.} \bibnamefont{Hu}}, \bibnamefont{and} \bibinfo{author}{\bibfnamefont{N.}~\bibnamefont{Ohta}}, \bibinfo{journal}{Phys. Rev. D} \textbf{\bibinfo{volume}{80}}, \bibinfo{pages}{104016} (\bibinfo{year}{2009}), \eprint{0910.2387}.

\bibitem[{\citenamefont{Hawking}(1975)}]{Hawking:1975vcx}
\bibinfo{author}{\bibfnamefont{S.~W.} \bibnamefont{Hawking}}, \bibinfo{journal}{Commun. Math. Phys.} \textbf{\bibinfo{volume}{43}}, \bibinfo{pages}{199} (\bibinfo{year}{1975}), \bibinfo{note}{[Erratum: Commun.Math.Phys. 46, 206 (1976)]}.

\bibitem[{\citenamefont{Izquierdo and Pavon}(2006)}]{Izquierdo:2005ku}
\bibinfo{author}{\bibfnamefont{G.}~\bibnamefont{Izquierdo}} \bibnamefont{and} \bibinfo{author}{\bibfnamefont{D.}~\bibnamefont{Pavon}}, \bibinfo{journal}{Phys. Lett. B} \textbf{\bibinfo{volume}{633}}, \bibinfo{pages}{420} (\bibinfo{year}{2006}), \eprint{astro-ph/0505601}.

\bibitem[{\citenamefont{Bekenstein}(1973)}]{Bekenstein:1973ur}
\bibinfo{author}{\bibfnamefont{J.~D.} \bibnamefont{Bekenstein}}, \bibinfo{journal}{Phys. Rev. D} \textbf{\bibinfo{volume}{7}}, \bibinfo{pages}{2333} (\bibinfo{year}{1973}).

\bibitem[{\citenamefont{Gong and Wang}(2007)}]{Gong:2007md}
\bibinfo{author}{\bibfnamefont{Y.}~\bibnamefont{Gong}} \bibnamefont{and} \bibinfo{author}{\bibfnamefont{A.}~\bibnamefont{Wang}}, \bibinfo{journal}{Phys. Rev. Lett.} \textbf{\bibinfo{volume}{99}}, \bibinfo{pages}{211301} (\bibinfo{year}{2007}), \eprint{0704.0793}.

\bibitem[{\citenamefont{Canetti et~al.}(2012)\citenamefont{Canetti, Drewes, and Shaposhnikov}}]{Canetti:2012zc}
\bibinfo{author}{\bibfnamefont{L.}~\bibnamefont{Canetti}}, \bibinfo{author}{\bibfnamefont{M.}~\bibnamefont{Drewes}}, \bibnamefont{and} \bibinfo{author}{\bibfnamefont{M.}~\bibnamefont{Shaposhnikov}}, \bibinfo{journal}{New J. Phys.} \textbf{\bibinfo{volume}{14}}, \bibinfo{pages}{095012} (\bibinfo{year}{2012}), \eprint{1204.4186}.

\bibitem[{\citenamefont{Kugo and Uehara}(1983)}]{Kugo:1982mr}
\bibinfo{author}{\bibfnamefont{T.}~\bibnamefont{Kugo}} \bibnamefont{and} \bibinfo{author}{\bibfnamefont{S.}~\bibnamefont{Uehara}}, \bibinfo{journal}{Nucl. Phys. B} \textbf{\bibinfo{volume}{222}}, \bibinfo{pages}{125} (\bibinfo{year}{1983}).

\bibitem[{\citenamefont{Bento et~al.}(2005)\citenamefont{Bento, Gonzalez~Felipe, and Santos}}]{Bento:2005xk}
\bibinfo{author}{\bibfnamefont{M.~C.} \bibnamefont{Bento}}, \bibinfo{author}{\bibfnamefont{R.}~\bibnamefont{Gonzalez~Felipe}}, \bibnamefont{and} \bibinfo{author}{\bibfnamefont{N.~M.~C.} \bibnamefont{Santos}}, \bibinfo{journal}{Phys. Rev. D} \textbf{\bibinfo{volume}{71}}, \bibinfo{pages}{123517} (\bibinfo{year}{2005}), \eprint{hep-ph/0504113}.

\bibitem[{\citenamefont{Sadjadi}(2007)}]{Sadjadi:2007dx}
\bibinfo{author}{\bibfnamefont{H.~M.} \bibnamefont{Sadjadi}}, \bibinfo{journal}{Phys. Rev. D} \textbf{\bibinfo{volume}{76}}, \bibinfo{pages}{123507} (\bibinfo{year}{2007}), \eprint{0709.0697}.

\bibitem[{\citenamefont{Luciano}(2026)}]{Luciano:2025tio}
\bibinfo{author}{\bibfnamefont{G.~G.} \bibnamefont{Luciano}}, \bibinfo{journal}{JHEAp} \textbf{\bibinfo{volume}{50}}, \bibinfo{pages}{100487} (\bibinfo{year}{2026}), \eprint{2510.00673}.

\bibitem[{\citenamefont{Tanabashi et~al.}(2018)}]{ParticleDataGroup:2018ovx}
\bibinfo{author}{\bibfnamefont{M.}~\bibnamefont{Tanabashi}} \bibnamefont{et~al.} (\bibinfo{collaboration}{Particle Data Group}), \bibinfo{journal}{Phys. Rev. D} \textbf{\bibinfo{volume}{98}}, \bibinfo{pages}{030001} (\bibinfo{year}{2018}).

\end{thebibliography}

\end{document}